\documentclass[letterpaper]{article} 
\usepackage{aaai25}  
\usepackage{times}  
\usepackage{helvet}  
\usepackage{courier}  
\usepackage[hyphens]{url}  
\usepackage{graphicx} 
\usepackage{natbib}  
\usepackage{caption} 
\usepackage{algorithm}
\usepackage{algpseudocode}

\usepackage{xcolor}
\usepackage{amsfonts}
\usepackage{amstext}
\usepackage{amsmath}
\usepackage{amsthm}
\usepackage{subfiles}
\usepackage{subfigure}
\usepackage{booktabs}
\usepackage{colortbl}
\usepackage{multirow}
\usepackage{bbding}
\usepackage{float}

\usepackage{bbding}
\usepackage{epsfig}
\usepackage{pifont}
\usepackage{amsthm}
\usepackage{amsmath}
\usepackage{amssymb}
\usepackage{subfiles}
\usepackage{multirow}
\usepackage{subfigure}

\newtheorem{theorem}{Theorem}
\newcommand{\tabincell}[2]{\begin{tabular}{@{}#1@{}}#2\end{tabular}}

\usepackage{newfloat}
\usepackage{listings}
\DeclareCaptionStyle{ruled}{labelfont=normalfont,labelsep=colon,strut=off} 
\floatstyle{ruled}
\newfloat{listing}{tb}{lst}{}
\floatname{listing}{Listing}
\title{Relation-Aware Equivariant Graph Networks for Epitope-Unknown \\ Antibody Design and Specificity Optimization}
\author{
Lirong Wu{$^{1,2,^*}$}, Haitao Lin{$^{1,2,^*}$}, Yufei Huang{$^{2}$}, Zhangyang Gao{$^{2}$}, Cheng Tan{$^{2}$}, \\ Yunfan Liu{$^{2}$}, Tailin Wu{$^{2}$}, Stan Z. Li{$^{2,^\dagger}$}\\
}
\affiliations{
$^1$ Zhejiang University, Hangzhou, China, 310058 \\
$^2$ AI Lab, Research Center for Industries of the Future, Westlake University, Hangzhou, China, 310030 \\
\{wulirong, linhaitao, huangyufei, gaozhangyang, tancheng, liuyunfan, wutailin, stan.zq.li\}@westlake.edu.cn}

\usepackage{bibentry}

\begin{document}

\maketitle

\begin{abstract}
Antibodies are Y-shaped proteins that protect the host by binding to specific antigens, and their binding is mainly determined by the Complementary Determining Regions (CDRs) in the antibody. Despite the great progress made in CDR design, existing computational methods still encounter several challenges: \textit{1)} poor capability of modeling complex CDRs with long sequences due to insufficient contextual information; \textit{2)} conditioned on pre-given antigenic epitopes and their \emph{static} interaction with the target antibody; \textit{3)} neglect of specificity during antibody optimization leads to non-specific antibodies. In this paper, we take into account a variety of node features, edge features, and edge relations to include more contextual and geometric information. We propose a novel \emph{\underline{R}elation-\underline{A}ware \underline{A}ntibody \underline{D}esign} (RAAD) framework, which dynamically models antigen-antibody interactions for co-designing the sequences and structures of antigen-specific CDRs. Furthermore, we propose a new evaluation metric to better measure antibody specificity and develop a contrasting specificity-enhancing constraint to optimize the specificity of antibodies. Extensive experiments have demonstrated the superior capability of RAAD in terms of antibody modeling, generation, and optimization across different CDR types, sequence lengths, pre-training strategies, and input contexts.
\end{abstract}

\section{Introduction}
Antibodies are a family of Y-shaped proteins produced by the immune system to recognize, bind, and neutralize pathogens (e.g., antigens)~\cite{basu2019recombinant}. Since the specificity of antibody-antigen binding is mainly determined by the Complementary Determining Regions (CDRs) in the antibody, how to design CDRs that bind to specific regions (epitopes) of an antigen becomes an important problem. Considering the enormous search space of over $20^{60}$ potential CDR sequence candidates and the high variability of CDR structures, it is not feasible to test all potential CDRs by experimental assays in the web laboratory, which calls for computational methods for antigen-binding antibody CDR design~\cite{tiller2015advances,kuroda2012computer}.

The computational antibody design has undergone a paradigm shift from energy-base models~\cite{adolf2018rosettaantibodydesign,lapidoth2015abdesign} to deep generative models~\cite{lin2024geoab,kong2022conditional,wu2024mape,wu2024learning,tan2024cross,jin2022antibody}. RefineGNN~\cite{jin2021iterative} is the first work to co-design CDR sequences and structures in an autoregressive manner, and its follow-up HSRN~\cite{jin2022antibody} unifies antigen-binding antibody design and docking in a hierarchical refinement framework. DiffAb~\cite{luo2022antigen} adopts diffusion models~\cite{hoogeboom2021argmax} to design the CDR sequences, $\mathrm{C}_\alpha$ coordinates, and orientations in an iterative manner. Similarly, AbDiffuser~\cite{martinkus2024abdiffuser} incorporates family-specific priors into the diffusion process, which greatly improves generation efficiency and quality over DiffAb. To improve computational efficiency, MEAN~\cite{kong2022conditional} replaces autoregressive decoding with progressive full-shot decoding and involves the light chain as a conditional input to generate CDRs. It is further extended as dyMEAN~\cite{kong2023end}, an end-to-end full-atom antibody design method. Two recent works, ABGNN~\cite{gao2023pre} and HTP~\cite{wu2023hierarchical}, focus on the benefits of antibody sequence pre-training and hierarchical pre-training, respectively. A detailed comparison of the characteristics of these methods has been summarized in Table.~A1 of \textbf{Appendix A}.

\begin{figure}[!tbp]
    \begin{center}
        \includegraphics[width=0.48\linewidth]{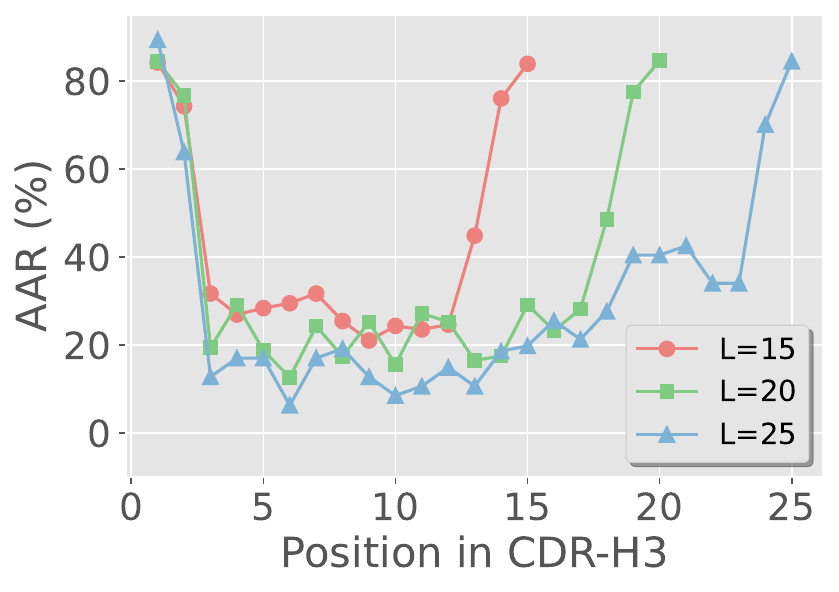}
	\includegraphics[width=0.48\linewidth]{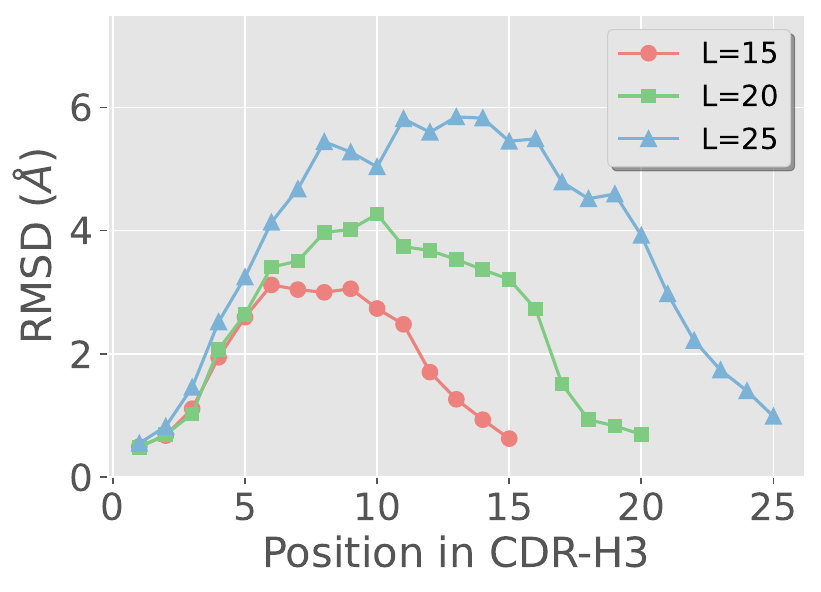}
    \end{center}
    \caption{Correlation of the performance of sequence generation (Left) and structure generation (Right) w.r.t amino acid position in CDR-H3 for different lengths $L$ from SAbDab.}
    \label{fig:1}
\end{figure}

Despite the fruitful progress, existing approaches still face several challenges. The \textbf{first} issue is that it is hard to model complex CDRs with long sequences. Take MEAN as an example in Fig.~\ref{fig:1}, it fails to model the types and coordinates of those amino acids in the middle of CDRs, especially for CDRs with long sequences. This is probably because the amino acids in the middle region of CDRs are far from the framework regions on either side and thus cannot acquire sufficient contextual information to make correct predictions. The \textbf{second} issue is that previous methods are mostly conditioned on pre-given epitopes~\cite{jin2022antibody,kong2022conditional} and take as input only the \emph{static} interactions between known epitopes and antibody. However, antigen-antibody interactions may evolve dynamically with the generation of different CDRs, especially when epitopes are not known. Note that dynamism is a polysemous term; it does not refer to dynamic conformations in this paper, but rather to the dynamism of antigen-antibody interaction graphs. The \textbf{third} issue is that existing antibody optimization methods may generate non-specific antibodies that not only bind with the target antigen but also other antigens with high affinity, thereby losing the important property of antibody specificity.

In this paper, we represent each antigen-antibody complex as an \textit{Attributed Heterogeneous Graph} (AHG) and define E(3)-invariant node and edge features in terms of six aspects: category, position, distance, direction, angle, and orientation, respectively, in order to include more geometrical information. Moreover, we defined a variety of different edge relations to expand the sequence and structural contexts of residues. We propose a simple but efficient \emph{\underline{R}elation-\underline{A}ware \underline{A}ntibody \underline{D}esign} (RAAD) framework that performs message passing for each relation separately and then jointly encodes different relational contexts into the final representations. RAAD takes the complete antigen (rather than known epitopes) as input and formulates antigen-antibody interaction dynamics modeling as an edge relation optimization problem. Furthermore, we propose a new evaluation metric for measuring antibody specificity by taking into account both the altered binding affinity of the optimized antibody to the target and non-target antigens. Moreover, a contrastive specificity-enhancing constraint is developed to improve the specificity of the optimized antibodies. Extensive experiments have demonstrated the superiority of RAAD in antibody modeling, generation, and optimization. Codes will be available at: \url{https://github.com/LirongWu/RAAD}.

\begin{figure}[!tbp]
    \begin{center}
	\includegraphics[width=1.0\linewidth]{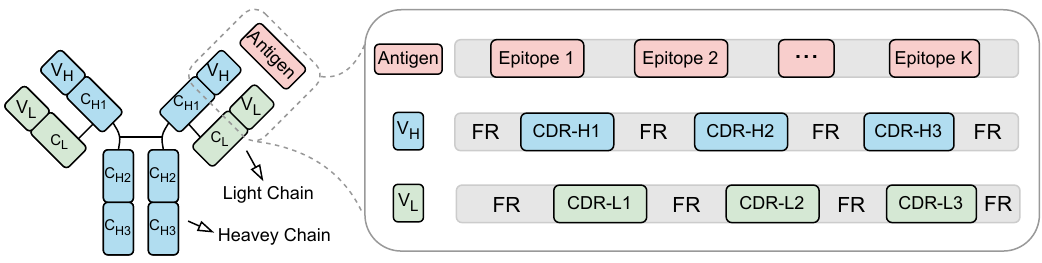}
    \end{center}
    \caption{Structure of an antigen-antibody complex, where the antibody is symmetrically Y-shape, and we mainly focus on three CDRs of the variable domain in the heavy chain.}
    \label{fig:2}
\end{figure}

\section{Preliminary}
An antibody is a $Y$-shaped symmetric protein with two identical sets of chains, as shown in Fig.~\ref{fig:2}. Each set contains a heavy chain and a light chain, either of which contains a variable domain ($V_H$/$V_L$) and several constant domains. Each variable domain can be further divided into four \emph{framework regions} (FRs) and three
\emph{complementarity-determining regions} (CDRs), each occupying a contiguous subsequence. The CDRs, especially those in the heavy chain (i.e., CDR-H1, CDR-H2, CDR-H3), are key in determining the antigen-antibody binding. Therefore, we mainly focus on the design and optimization of 3 CDRs in heavy chains in this paper.

\subsection{Graph Construction}
We represent each antigen-antibody complex as a heterogeneous graph $\mathcal{G}=(\mathcal{V}, \mathcal{E})$, where $\mathcal{V}=(\mathcal{V}_H, \mathcal{V}_L, \mathcal{V}_A)$ is the set of nodes (residues) of the heavy chain, the light chain and the antigen, and $\mathcal{E}$ contains the edges connecting the residues. Each node $v_i=(\mathbf{h}_i, \mathbf{X}_i)\in\mathcal{V}$ is attributed as a node feature vector $\mathbf{h}_i\in\mathbb{R}^{d_n}$ and a node coordinate matrix $\mathbf{X}_i\in\mathbb{R}^{3\times 4}$ consisting of 4 backbone atoms $\{N, C_{\alpha}, C, O\}$. In addition, each edge $e_{i,j}\in\mathcal{E}$ is described by an edge feature vector $\mathbf{E}_{i,j}\in\mathbb{R}^{d_e}$ and a set of edge relations $R_{i,j}\in\mathcal{R}$, where $\mathcal{R}$ contains 8 different relations describing the dependence between residue $v_i$ and residue $v_j$. Therefore, the graph of each antigen-antibody complex can also be denoted as $\mathcal{G}=(\mathbf{H}, \mathbf{X}, \mathbf{E}, R)$. To capture global information about different contextual components, we additionally insert a global node for the light chain, heavy chain, and antigen, respectively. The three global nodes are connected to each other and to all nodes within the same context component.

\textbf{Node and Edge Fetures} are defined in terms of 6 aspects: type, position, distance, direction, angle, and orientation. For each residue $v_i$, we define its node feature as follows
\begin{equation}
\hspace{-0.5em}
\begin{small}
\begin{aligned}
    \mathbf{h}_i&=\Big\{E_{\text{type}}(v_i), E_{\text{pos}}(i), \operatorname{sin}(\eta), \operatorname{cos}(\eta), \\  &\operatorname{RBF}(\|X_{i,C_\alpha}\!-\!X_{i,\xi}\|), Q_i^{\top}\frac{X_{i,\xi}\!-\!X_{i,C_\alpha}}{\left\|X_{i,\xi}\!-\!X_{i,C_\alpha}\right\|} \ \big|\  \eta, \xi\Big\},
\end{aligned}
\end{small}
\end{equation}
where $E_{\text{type}}(v_i)$ and $E_{\text{pos}}(i)$ are trainable type embedding and positional encoding of residue $v_i$, respectively. $\eta=\{\alpha_i,\beta_i,\gamma_i, \psi_i,\phi_i,\omega_i\}$ contains six bond/dihedral angles of residue $v_i$. Besides, $\operatorname{RBF}(\cdot)$ is a radial basis distance encoding function, and $X_{i,C_\alpha}$ is the coordinate of $C_\alpha$ atom in the $i$-th residue. The last term in $\mathbf{h}_i$ is a direction encoding that corresponds to the relative direction of three backbone atoms $\xi\in\{C_\beta,N,O\}$ in the local coordinate frame $Q_i$ of residue $v_i$. The edge feature $\mathbf{E}_{i,j}$ that describes the spatial relationship between two residues $v_i$ and $v_j$ is defined as follows
\begin{equation}
\hspace{-1.5em}
\begin{small}
\begin{aligned}
    \mathbf{E}_{i,j}\!=\!\Big\{E_{\text{type}}&(e_{i,j}), E_{\text{pos}}(i\!-\!j), \operatorname{RBF}(\|X_{i,C_\alpha}\!-\!X_{j,\xi}\|),  \\ & Q_i^{\top}\frac{X_{j,\xi}\!-\!X_{i,C_\alpha}}{\left\|X_{j,\xi}-X_{i,C_\alpha}\right\|}, q\left(Q_i^{\top} Q_j\right) \ \big| \ \xi\Big\},
\end{aligned}
\end{small}
\end{equation}
where $E_{\text{type}}(e_{i,j})$ is the one-encoding of relations $R_{i,j}$ between two residues, and the positional encoding $E_{\text{pos}}(i-j)$ encodes the relative sequential position. The third and fourth terms are distance and direction encodings of four backbone atoms $\xi\in\{C_\alpha, C_\beta,N,O\}$ in residue $v_j$ in the local coordinate frame $Q_i$. The last term $q\left(Q_i^{\top} Q_j\right)$ is the quaternion representation $q(\cdot)$ of $Q_i^{\top} Q_j$. Due to space limitations, we place a summary of node/edge features in \textbf{Appendix B}.

\begin{figure*}[!tbp]
	\begin{center}
	\includegraphics[width=1.0\linewidth]{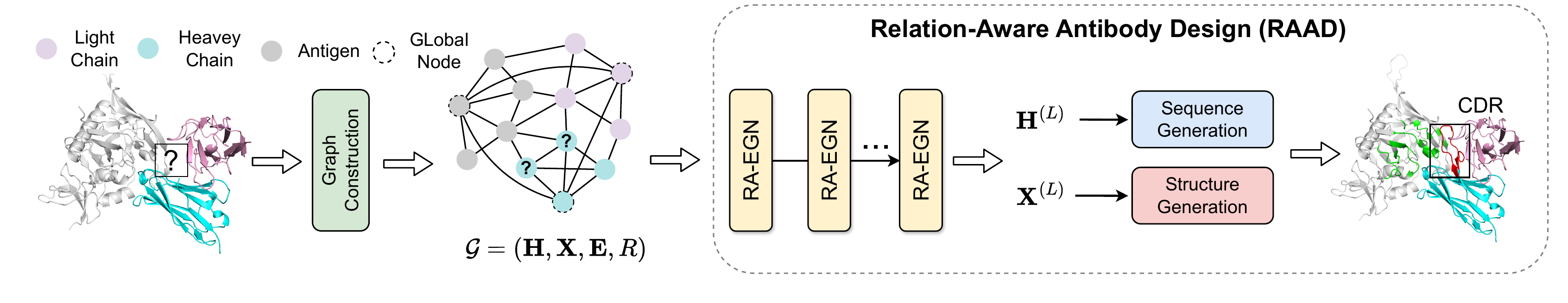}
	\end{center}
	\caption{A high-level overview of our RAAD framework. The antigen-antibody complex is constructed as an attributed heterogeneous graph, which is passed through multilayer Relation-Aware Equivariant Graph Networks (RA-EGN). The outputs are fed to sequence generation and structure predictor to generate both the sequence and structure of CDRs in the antibody.}
	\label{fig:3}
\end{figure*}

\textbf{Edge Relations.} To expand the contexts of CDRs, we divide the previous internal and external edges~\cite{kong2022conditional} into 8 different types of relations $\mathcal{R}=\{r_1,r_2,\cdots,r_8\}$, including \textit{(i)} \textbf{sequential relations} $r_1$ and $r_2$ between two residues with relative sequential distance equal to 1 and 2; \textit{(ii)} \textbf{internal spatial relations} between residues that are from the same component and spatially connected dut to $K$-nearest neighbors (relation $r_3$) or with a Euclidean distance less than 8 \textit{\AA} (relation $r_4$); \textit{(iii)} \textbf{external spatial relations} between residues that are from different components and spatially connected due to $K$-nearest neighbors (relation $r_5$) or with a Euclidean distance less than 12 \textit{\AA} (relation $r_6$); \textit{(iv)} \textbf{global relations}: relation $r_7$ between global nodes and relation $r_8$ between global nodes and nodes of the same component. Due to space limitations,  a detailed illustration of these edge relations is placed in \textbf{Appendix C}.

\subsection{Task Formulation}
The residue set of the target CDR to be generated is denoted as $\mathcal{V}_C\subseteq\mathcal{V}_H$ with node features and coordinates $\mathcal{C}=\{(\mathbf{h}_i,\mathbf{X}_i)\}_{v_i\in\mathcal{V}_C}$. Given a graph constructed from incomplete antigen-antibody complex $\mathcal{G} \backslash \mathcal{C}$, our objective is to learn a mapping $\mathcal{F}_{\Theta}: \mathcal{G}\backslash \mathcal{C} \rightarrow \{(s_i,\mathbf{X}_i)\}_{v_i\in\mathcal{V}_C}$, parameterized by $\Theta$, that generates the amino acid type $s_i$ and 3D coordinate $\mathbf{X}_i$ for each residue $v_i\in\mathcal{V}_C$ in the CDR.

\section{Relation-Aware Antibody Design (RAAD)}
The paper proposes a novel Relation-Aware Antibody Design (RAAD) framework, consisting of several layers of \textit{Relation-Aware Equivariant Graph Network} (RA-EGN). A high-level overview of our framework is shown in Fig.~\ref{fig:3}.

\subsection{Relation-Aware Equivariant Graph Network}
\begin{figure}[!htbp]
    \begin{center}
	\includegraphics[width=1.0\linewidth]{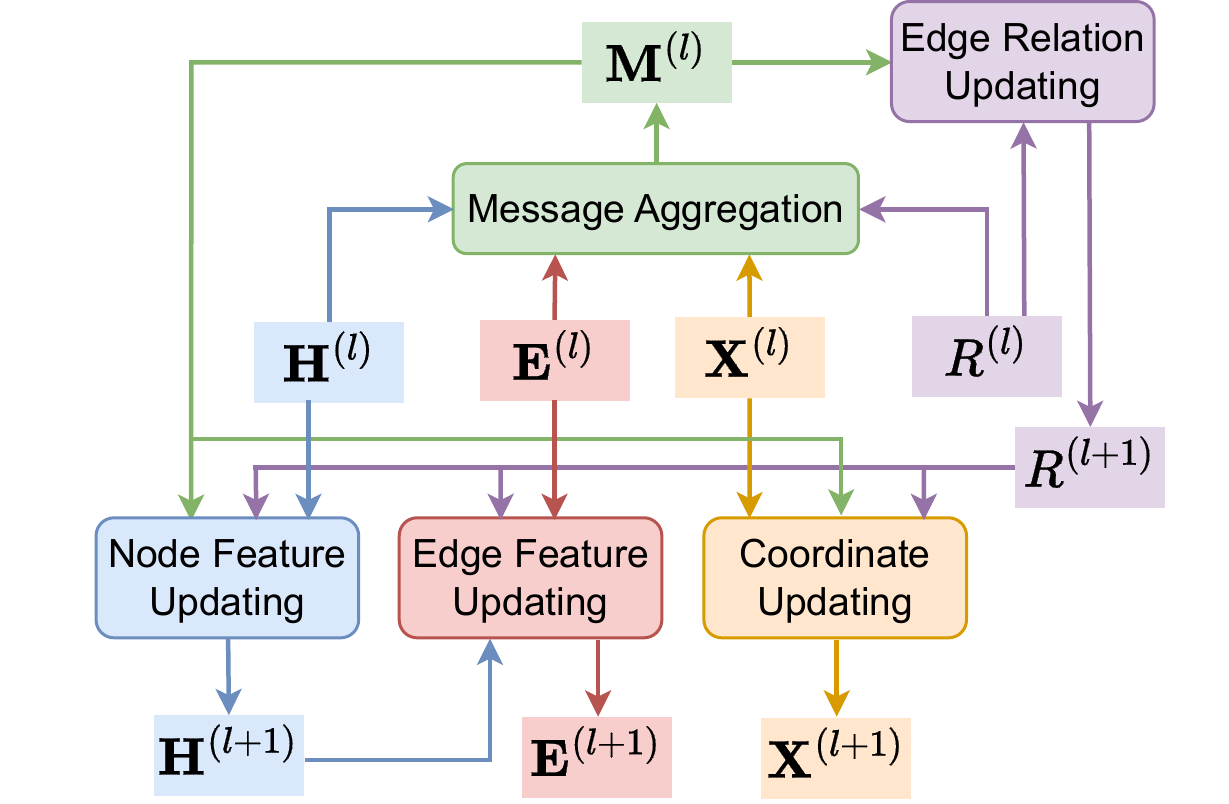}
    \end{center}
    	\caption{Each RA-EGN layer aggregates messages and jointly updates features, coordinates, and relations.}
    \label{fig:4}
\end{figure}

Each layer of RA-EGN takes into account the multi-scale residual interactions at the node, edge, and global levels and performs joint updating of node features, edge features, node coordinates, and edge relations to learn geometric residual representations. A schematic of RA-EGN is shown in Fig.~\ref{fig:4}.

\textbf{(1) Message Aggregation.} For any relation $r\in R_{i,j}$ that exist between residues $v_i$ and $v_j$, we perform relation-aware message aggregation with shared MLPs $\phi_m(\cdot)$, as follows
\begin{equation}
\mathbf{m}_{i,j,r}^{(l)}\!=\!\phi_m\Big(\mathbf{h}_i^{(l)}, \mathbf{h}_j^{(l)}, \frac{(\mathbf{X}_{i,j}^{(l)})^{\top} \mathbf{X}_{i,j}^{(l)}}{\big\|(\mathbf{X}_{i,j}^{(l)})^{\top} \mathbf{X}_{i,j}^{(l)}\big\|_F}, \mathbf{E}_{i,j}^{(l)}\Big),
\label{equ:3}
\end{equation}
where $\mathbf{h}_i^{(0)}=\mathbf{h}_i$, $\mathbf{X}_{i}^{(0)}=\mathbf{X}_{i}$, $\mathbf{E}_{i,j}^{(0)}=\mathbf{E}_{i,j}$, and $\mathbf{X}_{i,j}^{(l)}=\mathbf{X}_{i}^{(l)}\!-\!\mathbf{X}_{j}^{(l)}$ is the relative coordinates in the 3D space.

\begin{figure*}[!tbp]
	\begin{center}
		\includegraphics[width=1.0\linewidth]{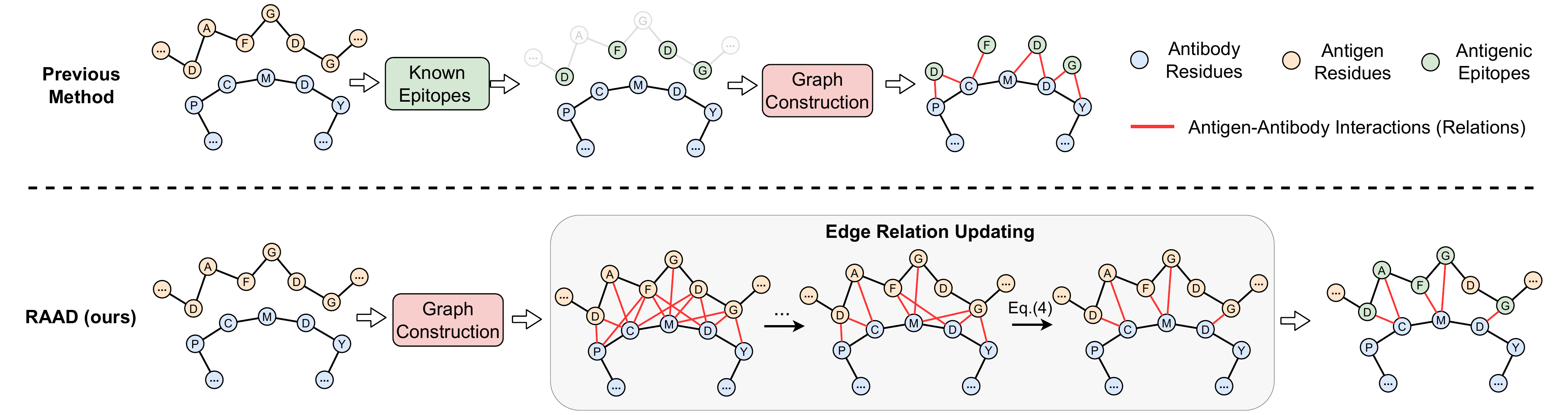}
	\end{center}
	\caption{A comparison between RAAD and previous methods in antigen-antibody interaction modeling. Their main differences include whether the interactions are dynamically updatable and whether antigenic epitopes are pre-given or adaptively learned.}
	\label{fig:5}
\end{figure*}

\textbf{(2) Edge Relation Updating.} Previous methods~\cite{kong2022conditional,kong2023end} mostly condition on pre-given epitopes and then consider only static interactions between known epitopes and antibodies. However, as different antibody CDRs are generated during training, the antigen-antibody interactions should also evolve dynamically. To achieve this, we transform antigen-antibody interaction dynamics modeling as an edge relation optimization problem. As shown in Fig.~\ref{fig:5}, RAAD takes dense antigen-antibody relations as inputs, then removes uninformative edge relations by edge relation updating, and keeps only those antigenic residues involved in antigen-antibody interactions as candidate epitopes. Compared to previous methods, \emph{it not only captures the dynamic properties of antigen-antibody interactions, but is also well suitable for epitope-unknown antibody design}. Specifically, the updating of edge relations focuses only on the relations $r_5$ and $r_6$ between the antigen and antibody (without those intra-component edge relations), and the whole updating process is defined as follows
\begin{equation}
\hspace{-0.5em}
\begin{aligned}
R^{(l+1)}_{i,j} = \Big\{r&\sim p_{i,j,r} \mid r\in M^{(l)}_{i,j}\Big\} \cup \Big(R^{(l)}_{i,j}\backslash M^{(l)}_{i,j}\Big), \\
\text{where}\ \  & p_{i,j,r} = \frac{\phi_z^{(r)}\big(\mathbf{m}_{i,j,r}^{(l)}\big)}{\max_{i,j} \phi_z^{(r)}\big(\mathbf{m}_{i,j,r}^{(l)}\big)},
\end{aligned}
\label{equ:4}
\end{equation}
where $M^{(l)}_{i,j} = \big\{r\mid r\in R^{(l)}_{i,j}, r=r_5 \text{ or } r_6, v_i \text{ or } v_j\in\mathcal{V}_A\big\}$ includes all relations between the antigen and antibody. Moreover, we model the interaction strength $\phi_z^{(r)}\big(\mathbf{m}_{i,j,r}^{(l)}\big)$ between residues $v_i$ and $v_j$ for each relation $r$ via MLPs $\phi_z^{(r)}(\cdot)$ and normalize it as the Bernoulli probability $p_{i,j,r}$. As a result, antigenic residues with higher interaction strengths will have a higher probability of being sampled.

\textbf{(3) Node Feature Updating.} We use the simplified graph transformer~\cite{gao2022pifold} to update
node representations, where we take node features $\mathbf{h}_i^{(l)}$, $\mathbf{h}_j^{(l)}$ and message $\mathbf{m}_{i,j,r}^{(l)}$ as inputs to learn the relation-aware attention weights $\alpha_{i,j,r}^{(l)}$ via MLPs $\phi_{\omega}^{(r)}(\cdot)$. We then aggregate messages for each relation separately based on the attention weights and merge them to update node features via MLPs $\phi_h(\cdot)$ as follows
\begin{equation}
\begin{aligned}
\mathbf{h}_i^{(l+1)}\!=\!\phi_h\Big(\mathbf{h}_i^{(l)}, &\sum_{r\in \mathcal{R}} \boldsymbol{W}_r^{(l)}\!\! \sum_{j \in \mathcal{N}_i^{(r)}} \alpha_{i,j,r}^{(l)} \ \mathbf{m}_{i,j,r}^{(l)}\Big), \\ \text{ where } \alpha_{i,j,r}^{(l)} &= \phi_{\omega}^{(r)}\Big(\mathbf{h}_i^{(l)}, \mathbf{m}_{i,j,r}^{(l)}, \mathbf{h}_j^{(l)}\Big),
\label{equ:5}
\end{aligned}
\end{equation}
where $\mathcal{N}_i^{(r)}$ is the neighborhood of residue $v_i$ that shares relation $r$ with $v_i$ in the updated relation set $R^{(l+1)}$. In addition, $\boldsymbol{W}_r^{(l)}$ is a kernel matrix shared by the relation $r\in\mathcal{R}$.

\textbf{(4) Edge Feature Updating.} The antigen-antibody complex is modeled as an attribute graph, which contains both node and edge features. However, previous works~\cite{jin2021iterative,kong2022conditional,kong2023end} only update node features and do not iteratively update edge features, which resulted in the dependencies between residues not being fully explored. We update the edge features in this paper based on the updated node features $\mathbf{h}_i^{(l+1)}$, $\mathbf{h}_j^{(l+1)}$ and the previous edge features $\mathbf{E}_{i,j}^{(l)}$ via MLPs $\phi_e(\cdot)$, as follows
\begin{equation}
\mathbf{E}_{i,j}^{(l+1)}=\phi_e\Big(\mathbf{h}_i^{(l+1)}, \mathbf{E}_{i,j}^{(l)}, \mathbf{h}_j^{(l+1)}\Big).
\end{equation}

\textbf{(5) Node Coordinate Updating.} Combining the interaction strength $\phi_z^{(r)}\big(\mathbf{m}_{i,j,r}^{(l)}\big)$ in Eq.~(\ref{equ:4}) and the attention weight $\alpha_{i,j,r}^{(l)}$ in Eq.~(\ref{equ:5}), we model the force fields of the residue $v_i$ with its surrounding neighbors for each relation $r$, and then update the coordinates in a one-shot manner, as follows
\begin{equation}
\begin{aligned}
\mathbf{X}_i^{(l+1)}\!=\!\mathbf{X}_i^{(l)}\!+\!\frac{1}{|\mathcal{R}|} \sum_{r\in\mathcal{R}} \sum_{j \in \mathcal{N}_i^{(r)}} \!\alpha_{i,j,r}^{(l)}\ \mathbf{X}_{i,j}^{(l)}\ \phi_z^{(r)}\big(\mathbf{m}_{i,j,r}^{(l)}\big).
\end{aligned}
\label{equ:7}
\end{equation}
We provide detailed derivations of feature and coordinate updating for global nodes in \textbf{Appendix D}. A desirable property of the proposed RA-EGN layer is E(3)-equivariance, as formally stated in Theorem~\ref{theorem:1}. Due to space limitations, a detailed proof of Theorem~\ref{theorem:1} can be found in \textbf{Appendix E}.
\begin{theorem}
We denote the $l$-th RA-EGN layer $g^{(l)}(\cdot)$ in our framework as $(\mathbf{H}^{(l+1)}, \mathbf{X}^{(l+1)}, \mathbf{E}^{(l+1)}, R^{(l+1)})=g^{(l)}(\mathbf{H}^{(l)}, \mathbf{X}^{(l)}, \mathbf{E}^{(l)}, R^{(l)})$. For any given orthogonal matrix $\mathbf{O}\in\mathbb{R}^{3\times 3}$ and translation matrix $\mathbf{t}\in\mathbb{R}^{3}$, $g^{(l)}(\cdot)$ is E(3)-equivariant, i.e., we have $(\mathbf{H}^{(l+1)}, \mathbf{O}\mathbf{X}^{(l+1)} + \mathbf{t}, \mathbf{E}^{(l+1)}, R^{(l+1)})=g^{(l)}(\mathbf{H}^{(l)}, \mathbf{O}\mathbf{X}^{(l)}+\mathbf{t}, \mathbf{E}^{(l)}, R^{(l)})$.
\label{theorem:1}
\end{theorem}

\subsection{CDR Sequence and Structure Generation}
Two common strategies for sequence and structure decoding are \textit{(1) autoregressive}~\cite{jin2021iterative,jin2022antibody}: generating CDR residues one by one; and \textit{(2) iterative refinement}~\cite{kong2022conditional,kong2023end}: the generation of the previous iteration is used as the initial for the next iteration. Autoregressive generation requires $N$ (length of the CDRs) times decoding, which suffers from the inefficiency problem. Although iterative refinement requires only $T \ll N$ iterations, it still suffers from error accumulation from previous iterations. However, our RAAD does not need either autoregression or iterative refinement, but rather directly generates both the sequences and structures of CDRs in the antibody \emph{in one single forward pass}.

\textbf{Sequence Generation.} The sequence generation module takes the encoded node representations $\mathbf{h}_i^{(L)}$ as input to predict the amino acid type, $\widehat{s}_i = \arg \max\big(\boldsymbol{W}_s\mathbf{h}_i^{(L)}\big)$,
where $\boldsymbol{W}_s$ is a linear transformation matrix. The loss function for sequence generation is defined as the cross-entropy $\ell_{ce}$ between the predicted amino acid types $\{\widehat{s}_i\}_{v_i\in\mathcal{V}_C}$ and the ground-truth types $\{s_i\}_{v_i\in\mathcal{V}_C}$, defined as follows
\begin{equation}
\mathcal{L}_{\text {seq }}=\frac{1}{|\mathcal{V}_C|} \sum_{v_i\in\mathcal{V}_C} \ell_{c e}\left(s_i, \operatorname{Softmax}\big(\boldsymbol{W}_s\mathbf{h}_i^{(L)}\big)\right)
\label{equ:8}
\end{equation}
\textbf{Structure Generation.} We directly take the node coordinate $\mathbf{X}_i^{(L)}$ from the last layer of RA-EGN as the backbone structure $\widehat{\mathbf{X}}_i=\mathbf{X}_i^{(L)}$ without using any additional decoding process. Furthermore, we can perform the side-chain packing using Rosetta~\cite{chaudhury2010pyrosetta} to obtain the full-atom 3D structure. As for the loss function for structure generation, we adopt the Huber loss $\ell_{\text {huber }}(\cdot,\cdot)$~\cite{huber1992robust} (please refer to \textbf{Appendix F} for detailed formulas) other than the common MSE, defined as follows
\begin{equation}
\mathcal{L}_{\text {struct }}=\frac{1}{\left|\mathcal{V}_C\right|} \sum_{v_i \in \mathcal{V}_C} \ell_{\text {huber }}\left(\widehat{\mathbf{X}}_i, \mathbf{X}_i\right).
\label{equ:9}
\end{equation}

\subsection{Antibody Specificity Optimization}
Antibody optimization is another important aspect of antibody design besides antibody generation, and it is aimed towards antibodies that are not only structurally realistic but also have desirable properties, such as specificity. Most of the existing work~\cite{kong2022conditional,gao2023pre,tan2024cross,kong2023end} adopts the change in binding energy ($\Delta\Delta G$, kcal/mol) predicted by a geometric network~\cite{shan2022deep} to measure antibody-antigen binding affinity, but it does not reflect the specificity of the antibody. To address this problem, we define a new evaluation metric to measure antibody specificity by considering $\Delta\Delta G$s of the optimized antibody to both target and non-target antigens in the test set $\mathcal{D}$, defined as follows
\begin{equation}
\begin{aligned}
    \text{SP-score}& = \Delta\Delta G_{\text{neg}} - \Delta\Delta G_{\text{pos}}, \ \text{where} \\
     \Delta\Delta G_{\text{pos}}&= \frac{1}{|\mathcal{D}|}\sum_{i\in\mathcal{D}} \Delta\Delta G_{i,i} \\
    \Delta\Delta G_{\text{neg}} &= \frac{1}{|\mathcal{D}|}\sum_{i\in\mathcal{D}}\Big(\frac{1}{|\mathcal{D}|-1}\sum_{j\in\mathcal{D}\backslash i} \Delta\Delta G_{i,j}\Big),
\end{aligned}
\end{equation}
where $\Delta\Delta G_{i,j}$ denotes the $\Delta\Delta G$ score between $i$-th antibody and $j$-th antigen. A larger $\text{SP-score}$ indicates that the antibody has better specificity. To perform antibody optimization, previous approaches have directly applied \emph{Iterative Target Augmentation} (ITA)~\cite{yang2020improving} to generate high-quality antibodies as training data to fine-tune the antibody generator $\mathcal{F}_{\Theta}$. However, we found that such antibody optimization tends to produce ``universal antibodies" that not only bind to the target antigen but also other antigens with high affinity, thereby losing the important property of antibody specificity. To improve the antibody specificity during ITA-based antibody optimization, we propose in this paper a contrastive specificity-enhancing constraint $\mathcal{L}_{\text{SP}}$ based on mutual information, as follows
\begin{equation}
\begin{aligned}
    \mathcal{L}_{\text{SP}} = \mathbb{E}_{\mathbb{P}_{\mathcal{C}_i}}\Big\{&-\text{sim}(f_1(\mathcal{C}_i), f_2(\mathcal{A}_i)) \\ & +\log\big(\mathbb{E}_{\mathbb{P}_{\mathcal{A}_j}}e^{\text{sim}\left(f_1(\mathcal{C}_i), f_2(\mathcal{A}_j)\right)}\big)\Big\},
\end{aligned}
\label{equ:11}
\end{equation}
where $f_1(\mathcal{C}_i)=\frac{1}{|\mathcal{V}_C|}\sum_{v_j\in\mathcal{V}_C}\mathbf{h}_j^{(L)}$ is the embedding of CDR $\mathcal{C}_i$ in the $i$-th antibody, $f_2(\mathcal{A}_i)=\frac{1}{|\mathcal{V}_A|}\sum_{v_j\in\mathcal{V}_A}\mathbf{h}_j^{(L)}$ is the averaged embedding of residues in the $i$-th antigen $\mathcal{A}_i$, and $\text{sim}(\cdot,\cdot)$ denotes the cosine similarity. The loss $\mathcal{L}_{\text{SP}}$ esentially maximizes a lower bound of the mutual information~\cite{you2020graph} between $f_1(\mathcal{C}_i)$ and $f_2(\mathcal{A}_i)$ ,i.e., the corresponding antibody and antigen. Following \citep{jin2021iterative}, we exploit the Iterative Target Augmentation (ITA) \cite{yang2020improving} algorithm to tackle the antibody optimization problem. Specifically, we maintain a high-affinity candidate queue for each antibody to optimize the distribution of generated antibodies towards the high-affinity landscape. In each ITA step, we first generate 50 candidates for each antibody and sort them together with the candidates in the candidate queue according to their predicted $\Delta\Delta G$ scores. Then, only the four candidate antibodies with the highest affinity scores will be kept in the queue and used to fine-tune the antibody generator $\mathcal{F}_{\Theta}$ by minimizing the joint loss of constraint $\mathcal{L}_{\text{SP}}$ and losses $\mathcal{L}_{\text{seq}}$, $\mathcal{L}_{\text{struct}}$, as follows
\begin{equation}
    \mathcal{L}_{\text{opt}} = (\mathcal{L}_{\text{seq}} + \eta\mathcal{L}_{\text{struct}}) + \mathcal{L}_{\text{SP}},
\label{equ:12}
\end{equation}
where $\eta$ is a hyperparameter that controls the loss balance. The pseudo-code of ITA-based antibody optimization is placed in \textbf{Appendix G}. Due to space limitations, the time complexity analysis of RAAD is available in \textbf{Appendix H}.

\section{Experiments}
Extensive experiments are conducted on the six aspects: (1) Sequence and structure modeling; (2) Antigen-binding CDR-H3 generation; (3) Antibody optimization; (4) Ablation study; (5) Evaluation of different CDR sequence lengths and input contexts; (6) Evaluation on pre-training strategies.

\textbf{Baselines.} We compare RAAD with several state-of-the-art methods, include LSTM~\cite{saka2021antibody}, RefineGNN~\cite{jin2021iterative}, HERN~\cite{jin2022antibody}, DiffAb~\cite{luo2022antigen}, AbDiffuser~\cite{martinkus2024abdiffuser}, ADesigner~\cite{tan2024cross}, MEAN~\cite{kong2022conditional}, and dyMEAN~\cite{kong2023end}. Since LSTM, RefineGNN, and HERN encode only partial contexts of the antigen-antibody complex, either heavy chain (LSTM and RefineGNN) or antigen (HERN), we consider three variants, C-LSTM, C-RefineGNN, and C-HERN, by taking the complete antigen-antibody complex as the input context. The detailed implementations and hyperparameters are placed in \textbf{Appendix I}. 

\begin{figure*}[!tbp]
	\begin{center}
	\includegraphics[width=0.75\linewidth]{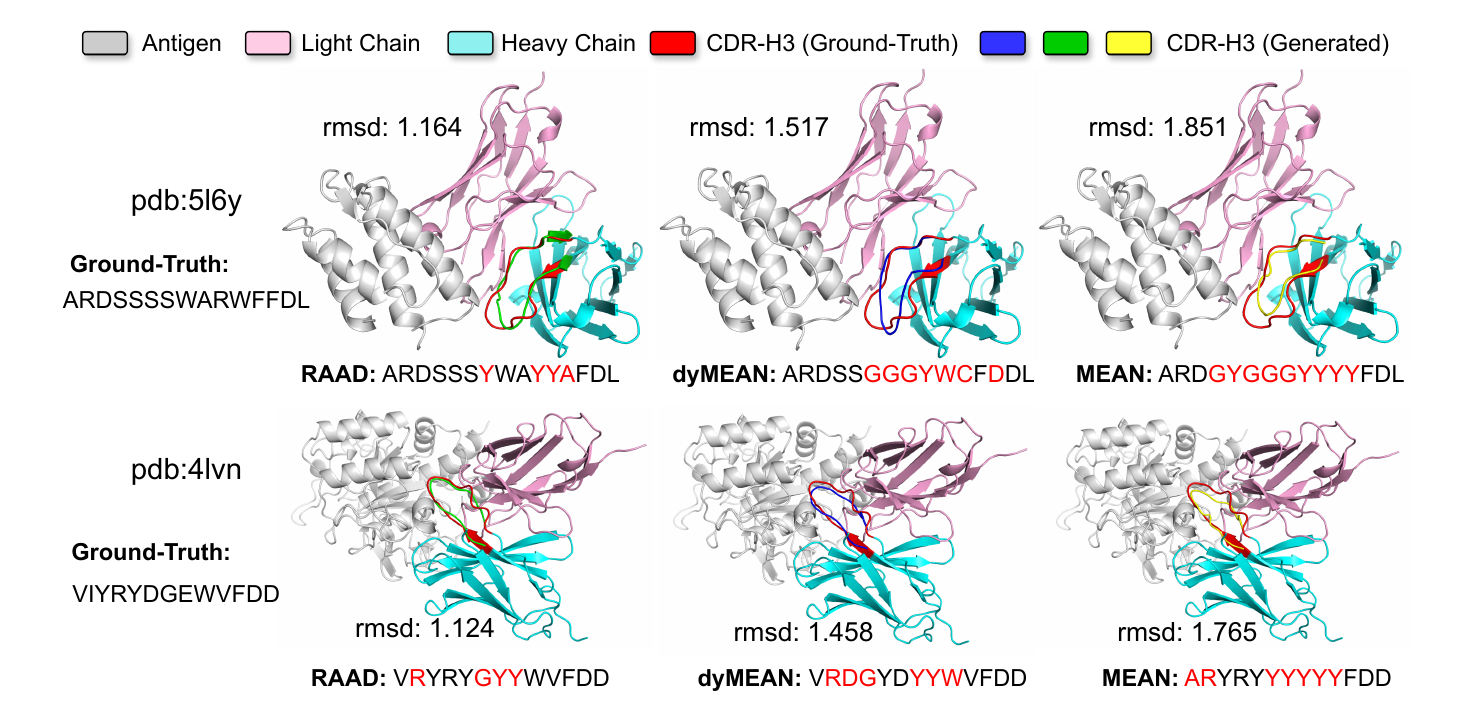}
	\end{center}
	\caption{Case Study. The structures and sequences of the antigen-binding CDR-H3s designed by MEAN, dyMEAN, and RAAD for two complexes (pdb: 5l6y and 4lvn), as well as RMSD scores with the ground-truth CDR-H3 structures.}
	\label{fig:6}
\end{figure*}

\subsection{Sequence and Structure Modeling}
We evaluate the generative modeling capacity of RAAD in the Structural Antibody Database (SAbDab)~\cite{lai2018fast}, which contains 3,127 antigen-antibody complexes after filtering structures that lack light chains or antigens. We first cluster the CDR sequences via MMseqs2~\cite{steinegger2017mmseqs2} that assigns the antibodies with CDR sequence identity~\cite{henikoff1992amino} above 40\% to the same cluster~\cite{kong2022conditional}. After preprocessing, the numbers of clusters for CDR-H1, CDR-H2, and CDR-H3 are 765, 1093, 1659, respectively, which are divided into the training, validation, and test sets with a split ratio of 8: 1: 1. We employ Amino Acid Recovery (AAR) and Root Mean Squared Deviation (RMSD) as evaluation metrics for sequence generation and structure prediction, respectively, and report the results of 10-fold cross-validation in Table.~\ref{tab:1}. For two diffusion-based generative models, AiffAb and AbDiffuser, we generate 100 candidate samples for each CDR and finally report the best ones. It can be found that RAAD outperforms previous baselines on all three CDR regions, especially in the CDR-H2 and for the task of structure prediction. For example, RAAD surpasses dyMEAN by 6.24\% in AAR and 12.20\% in RMSD in the CDR-H2. 

\begin{table}[!tbp]
\begin{center}
\caption{Mean results of 10-fold cross-validation for CDR sequence and structure modeling on the SAbDab, where \textbf{bold} and \underline{underline} denote the best and second metrics.}
\label{tab:1}
\resizebox{1.0\columnwidth}{!}{
\begin{tabular}{l|cc|cc|cc}

\toprule
\multirow{2}{*}{\textbf{Model}} & \multicolumn{2}{c|}{\textbf{CDR-H1}} & \multicolumn{2}{c|}{\textbf{CDR-H2}} & \multicolumn{2}{c}{\textbf{CDR-H3}} \\ \cmidrule(r){2-3} \cmidrule(r){4-5} \cmidrule(r){6-7}
 & \ \ AAR ($\uparrow$)\ \  & RMSD ($\downarrow$) & \ \ AAR ($\uparrow$)\ \  & RMSD ($\downarrow$) & \ \ AAR ($\uparrow$)\ \  & RMSD ($\downarrow$) \\ \midrule
LSTM & 49.98 & - & 28.50 & - & 15.69 & - \\
C-LSTM & 40.93 & - & 29.24 & - & 15.48 & - \\
RefineGNN & 39.40 & 3.22 & 37.06 & 3.64 & 21.13 & 6.00 \\
C-RefineGNN\quad\quad & 33.19 & 3.25 & 33.53 & 3.69 & 18.88 & 6.22 \\
HERN & 41.71 & 1.78 & 35.46 & 2.21 & 24.75 & 3.89 \\
C-HERN & 47.36 & 1.45 & 41.45 & 1.20 & 28.47 & 2.64 \\ \midrule
DiffAb & 61.34 & 1.02 & 37.66 & 1.20 & 25.79 & 3.02 \\
AbDiffuser & 62.76 & 1.08 & 41.10 & 1.16 & 29.58 & 2.83 \\
MEAN & 58.29 & 0.98 & 47.15 & 0.95 & 36.38 & 2.21 \\
dyMEAN & 63.52 & 0.84 & 55.41 & 0.82 & 37.19 & 2.09 \\
ADesigner & \underline{64.34} & \underline{0.82} & \underline{55.52} & \underline{0.79} & \underline{37.37} & \underline{1.97} \\
RADD (ours) & \textbf{67.64} & \textbf{0.77} & \textbf{58.87} & \textbf{0.72} & \textbf{37.71} & \textbf{1.91} \\ \midrule
$\Delta_{\text{dyMEAN}}$ & +6.49\% & -8.33\% & +6.24\% & -12.20\% & +1.67\% & -8.61\% \\ \bottomrule

\end{tabular}}
\end{center}
\end{table} 

\begin{table}[!tbp]
\begin{center}
\caption{Performance comparison of antigen-binding CDR-H3 sequence and structure design on the RAbD dataset.}
\label{tab:2}
\resizebox{0.8\columnwidth}{!}{
\begin{tabular}{l|ccc}

\toprule
        \textbf{Model} & AAR ($\uparrow$) & TM-score ($\uparrow$) & RMSD ($\downarrow$) \\ \midrule
        RosettaAD & 22.50 & 0.9435 & 5.52 \\ \midrule
        LSTM & 22.36 & - & - \\
        C-LSTM & 22.18 & - & - \\
        RefineGNN & 29.79 & 0.8308 & 7.55 \\
        C-RefineGNN \quad\quad & 28.90 & 0.8317 & 7.21 \\
        HERN & 32.83 & 0.9684 & 3.12 \\
        C-HERN & 34.51 & 0.9734 & 2.79 \\
        MEAN & 36.77 & 0.9812 & 1.81 \\
        dyMEAN & 39.14 & 0.9825 & 1.66 \\
        ADesigner & \underline{40.94} & \underline{0.9850} & \underline{1.55} \\
        RAAD (ours) & \textbf{41.26} & \textbf{0.9874} & \textbf{1.46} \\ \midrule
        $\Delta_{\text{dyMEAN}}$ & +5.42\% & +0.50\% & -12.05\% \\\bottomrule

\end{tabular}} 
\end{center}
\end{table} 

\subsection{Antigen-binding CDR-H3 Generation}
We evaluate how well RAAD designs CDR-H3 regions that bind to specific antigens on the RAbD dataset~\cite{adolf2018rosettaantibodydesign}, which contains 60 complexes. We still use the SAbDab dataset as the training data, but remove the antibodies in which their CDR-H3 has the same cluster as RAbD, and split it into training and validation sets with a split ratio of 9:1. In addition to AAR and RMSD, we also adopt TM-score as an additional metric for calculating the global similarity between the predicted and ground-truth structures. As can be seen from Table.~\ref{tab:2}, our RAAD achieves the best results across three metrics, suggesting its superiority in designing antigen-binding CDR-H3 regions. 

To evaluate the generative capacity of RAAD, we visualize the results of CDR-H3 generation for two complexes (pdb: 5l6y and 4lvn) in Fig.~\ref{fig:6} along with ground-truth and predicted structure prediction errors (RMSD scores). It can be seen that the accuracy of RAAD sequence and structure generation is much higher than that of MEAN and DyMEAN. For example, the RMSD scores of CDR-H3 structures (pdb: 5l6y) designed by MEAN, DyMEAN and RAAD are 1.851 \textit{\AA}, 1.517 \textit{\AA} and 1.164 \textit{\AA}, respectively.

\subsection{Antibody Specificity Optimization}
Antibody optimization towards higher binding affinity and specificity is an important goal of antibody design. To evaluate the effectiveness of RAAD in antibody optimization, we first pre-train the antibody generator $\mathcal{F}_{\Theta}$ on the SAbDab dataset with a split ratio of 9:1 for training and validation. Next, we fine-tune the antibody generator $\mathcal{F}_{\Theta_*}$ in the SKEMPI V2.0 dataset containing 53 antigen-antibody complexes for antibody optimization. Table.~\ref{tab:3} reports the average change in binding affinity of antibodies to the target antigen and other antigens, $\Delta\Delta G_{\text{pos}}$ and $\Delta\Delta G_{\text{neg}}$, as well as their differences, i.e., the specificity SP-scores. Besides, we also report the number of changed residues $\Delta L$ since many practical scenarios
prefer smaller $\Delta L$~\cite{ren2022proximal}. Previous works focused only on $\Delta\Delta G_{\text{pos}}$ in antibody optimization, and RAAD achieved the best in this metric with minimal $\Delta L$. Furthermore, we found that $\mathcal{L}_{\text{SP}}$ defined in Eq.~(\ref{equ:11}) can significantly improve $\Delta\Delta G_{\text{neg}}$, leading to better specificity metrics, i.e., SP-score. For example, $\Delta\Delta G_{\text{neg}}$ of RAAD with $\mathcal{L}_{\text{SP}}$ is only -0.238, compared to -1.605 of MEAN and -1.543 of dyMEAN, suggesting that $\mathcal{L}_{\text{SP}}$ helps to improve the \emph{binding specificity} of optimized antibodies. 




\begin{table}[!tbp]
\begin{center}
\caption{Average $\Delta\Delta G$ (kcal/mol) and average number of changed residues ($\Delta L$) after antibody optimization.}
\label{tab:3}
\resizebox{1.0\columnwidth}{!}{
\begin{tabular}{l|cccc}

\toprule
\textbf{Model} & $\Delta\Delta G_{\text{pos}}$ ($\downarrow$) & $\Delta\Delta G_{\text{neg}}$ ($\uparrow$) & SP-score ($\uparrow$) & $\Delta L$ ($\downarrow$) \\ \midrule
Random & +1.489 & \textbf{+1.518} & 0.029 & 15.64 \\
LSTM & -2.183 & -0.496 & 1.687 & 9.45 \\
C-LSTM & -2.534 & -0.784 & 1.750 & 10.24 \\
RefineGNN & -4.684 & -1.191 & 3.493 & 7.76 \\
C-RefineGNN & -5.179 & -1.392 & 3.787 & 8.56 \\
MEAN & -6.104 & -1.605 & 4.499 & 9.61 \\
dyMEAN & -8.215 & -1.543 & 6.672 & \underline{6.24} \\
ADesigner & -10.431 & -1.624 & 8.807 & 6.82 \\ \midrule
RAAD (w/o $\mathcal{L}_{\text{SP}}$) & \textbf{-12.163} & -1.748 & \underline{10.415} & 6.50 \\
RAAD (w/ $\mathcal{L}_{\text{SP}}$) & \underline{-11.787} & \underline{-0.238} & \textbf{11.549} & \textbf{5.98} \\ \bottomrule

\end{tabular}} 

\end{center} 
\end{table}

\begin{figure}[!hbp]
	\begin{center}
	\includegraphics[width=0.8\linewidth]{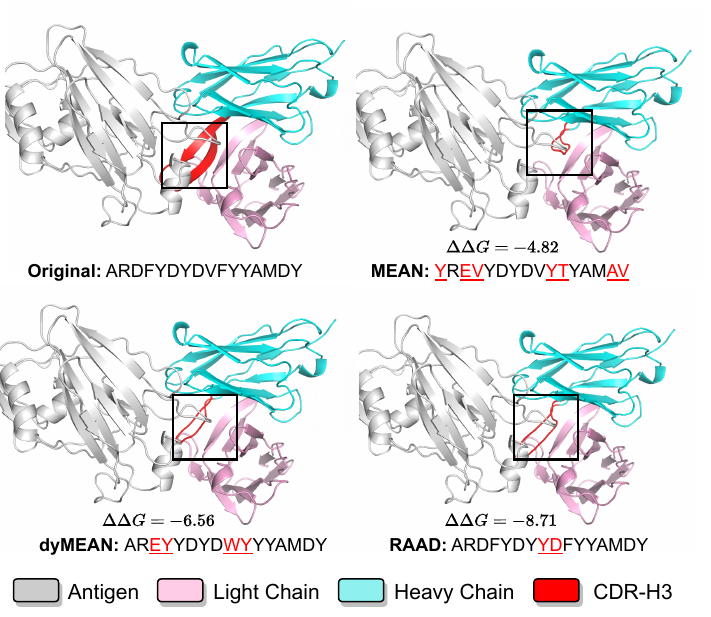}
	\end{center}
	\caption{3D structures and $\Delta\Delta G$ of CDR-H3s (pdb: 2vis) optimized by MEAN, dyMEAN, RAAD, as well as the original and optimized sequences (altered are marked in red).}
	\label{fig:7}
\end{figure}

Lastly, we visualize an example of the optimized CDH-H3 by MEAN, dyMEAN, and RAAD in Fig.~\ref{fig:7}, including the optimized CDR sequences and structures, as well as $\Delta\Delta G$s. It shows that RAAD maximally improves the binding affinity ($\Delta\Delta G\!=\!-8.17$) with minimal mutations ($\Delta L\!=\!2$).

\subsection{Ablation Study}
We further conduct ablation study on five key components: (1) \emph{w/o node features}, using only node type features $E_{\text{type}}(v_i)$; (1) \emph{w/o edge features}, using only edge type features $E_{\text{type}}(e_{i,j})$; (3) \emph{w/o relational MMP}, which performs message passing only once over all edges; (4) \emph{w/o dynamics modeling}, no longer dynamically updating edge relations across layers, but directly using the initial (static) antigen-antibody relations instead; (5) \emph{w/o global nodes}, removing all global nodes and the involved message passing. From the results in Table.~\ref{tab:4}, it can be seen that relation-aware MMP and interaction dynamics modeling play the most important roles. In addition, both node and edge features help to enrich the contextual information of the residue and thus improve the performance. Furthermore, we find that the removal of global nodes results in a significant performance drop, which illustrates the importance of global message passing between different input context components.

\begin{table}[!tbp]
\begin{center}
\caption{Ablation study on different model componens, where \textbf{bold} and {\color[rgb]{0.25, 0.25, 0.95}gray} denote the best and poorest metrics.}
\label{tab:4}
\resizebox{\columnwidth}{!}{
\begin{tabular}{l|cccccc}

\toprule
\multirow{2}{*}{\textbf{Model}} & \multicolumn{2}{c}{\textbf{Modeling}} & \multicolumn{3}{c}{\textbf{Generation}} & \textbf{Optimization} \\ \cmidrule(r){2-3} \cmidrule(r){4-6} \cmidrule(r){7-7} 
 & AAR$\uparrow$ & RMSD$\downarrow$ & AAR$\uparrow$ & TM-score$\uparrow$ & RMSD$\downarrow$ & SP-score$\uparrow$ \\
 \midrule
full model & \textbf{67.64} & \textbf{0.772} & \textbf{39.86} & \textbf{0.9847} & \textbf{1.442} & \textbf{11.55} \\
\ - node feature & 66.12 & 0.794 & 38.49 & 0.9829 & 1.522 & 10.12 \\
\ - edge feature & 65.62 & {\color[rgb]{0.25, 0.25, 0.95}0.819} & 38.66 & 0.9825 & 1.517 & 9.75 \\
\ - relational MMP & {\color[rgb]{0.25, 0.25, 0.95}64.85} & 0.807 & {\color[rgb]{0.25, 0.25, 0.95}37.97} & {\color[rgb]{0.25, 0.25, 0.95}0.9804} & {\color[rgb]{0.25, 0.25, 0.95}1.596} & {\color[rgb]{0.25, 0.25, 0.95}8.94} \\
\ - edge updating & 66.73 & 0.798 & 39.14 & 0.9836 & 1.488 & 10.74 \\
\ - global nodes & 65.63 & 0.810 & 38.27 & 0.9817 & 1.549 & 9.60 \\ \bottomrule

\end{tabular}} 
\end{center}
\end{table}

\subsection{Evaluation on Length, Context, and Pre-training}

\textbf{Input Context.} To evaluate how well RAAD is applicable to different input contexts, we consider five contextual combinations as in Table.~\ref{tab:5}. In particular, we explore how well RAAD performs if the known epitopes are pre-given and directly taken as inputs instead of the full antigen~\cite{kong2022conditional,kong2023end}. It can be observed from the results in Table.~\ref{tab:5} that (1) contexts on heavy chains, i.e., framework regions, play the most important role; (2) light chain helps little in antibody design; and (3) dynamic modeling of interactions between the full antigen and antibody can greatly improve performance across a variety of tasks compared to considering only those pre-given epitopes.

\begin{table}[!htbp]
\begin{center}
\caption{Evaluation on different input contexts, ranging from light chains, heavy chains, full antigens, and epitopes. The performance gains and drops are marked as {\color[rgb]{0.8, 0.227, 0.141}red} and {\color[rgb]{0.4, 0.71, 0.376}green}.}
\label{tab:5}
\resizebox{1.0\columnwidth}{!}{
\begin{tabular}{c|ccc|ccccc}

\toprule
\multirow{2}{*}{} & \multicolumn{3}{c|}{\textbf{Model}} & \multicolumn{2}{c}{\textbf{Modeling}} & \multicolumn{3}{c}{\textbf{Generation}} \\ \cmidrule(r){2-4} \cmidrule(r){5-6} \cmidrule(r){7-9} 
& Heavey & Light & Antigen & AAR ($\uparrow$) & RMSD ($\downarrow$) & AAR ($\uparrow$) & TM-score ($\uparrow$) & RMSD ($\downarrow$) \\ \midrule
\Large\ding{172} & \XSolidBrush & \XSolidBrush & Full & 54.89 & 0.943 & 36.56 & 0.9795 & 1.818 \\
\Large\ding{173} & \XSolidBrush & \Checkmark & Full & 58.56 & 0.915 & 37.74 & 0.9783 & 1.774 \\
\Large\ding{174} & \Checkmark & \XSolidBrush & Full & 66.05 & 0.814 & \underline{40.17} & \underline{0.9862} & \textbf{1.442} \\
\Large\ding{175} & \Checkmark & \Checkmark & Epitope & \underline{66.54} & \underline{0.796} & 39.42 & 0.9839 & 1.521 \\
\Large\ding{176} & \Checkmark & \Checkmark & Full & \textbf{67.64} & \textbf{0.772} & \textbf{41.26} & \textbf{0.9874} & \underline{1.463} \\ \midrule

\multicolumn{4}{c|}{\Large(\Large\ding{176} - \Large\ding{174}\Large) / \Large\ding{174}} & {\color[rgb]{0.8, 0.227, 0.141}+2.31\%} & {\color[rgb]{0.8, 0.227, 0.141}-5.16\%} & {\color[rgb]{0.8, 0.227, 0.141}+2.71\%} & {\color[rgb]{0.8, 0.227, 0.141}+0.21\%} & {\color[rgb]{0.4, 0.71, 0.376}+1.46\%} \\ 
\multicolumn{4}{c|}{\Large(\Large\ding{176} - \Large\ding{175}\Large) / \Large\ding{175}} & {\color[rgb]{0.8, 0.227, 0.141}+1.65\%} & {\color[rgb]{0.8, 0.227, 0.141}-3.02\%} & {\color[rgb]{0.8, 0.227, 0.141}+4.67\%} & {\color[rgb]{0.8, 0.227, 0.141}+0.36\%} & {\color[rgb]{0.8, 0.227, 0.141}-3.81\%} \\ \bottomrule

\end{tabular}}
\end{center}
\end{table}

\begin{figure}[!tbp]
	\begin{center}
		\includegraphics[width=0.48\linewidth]{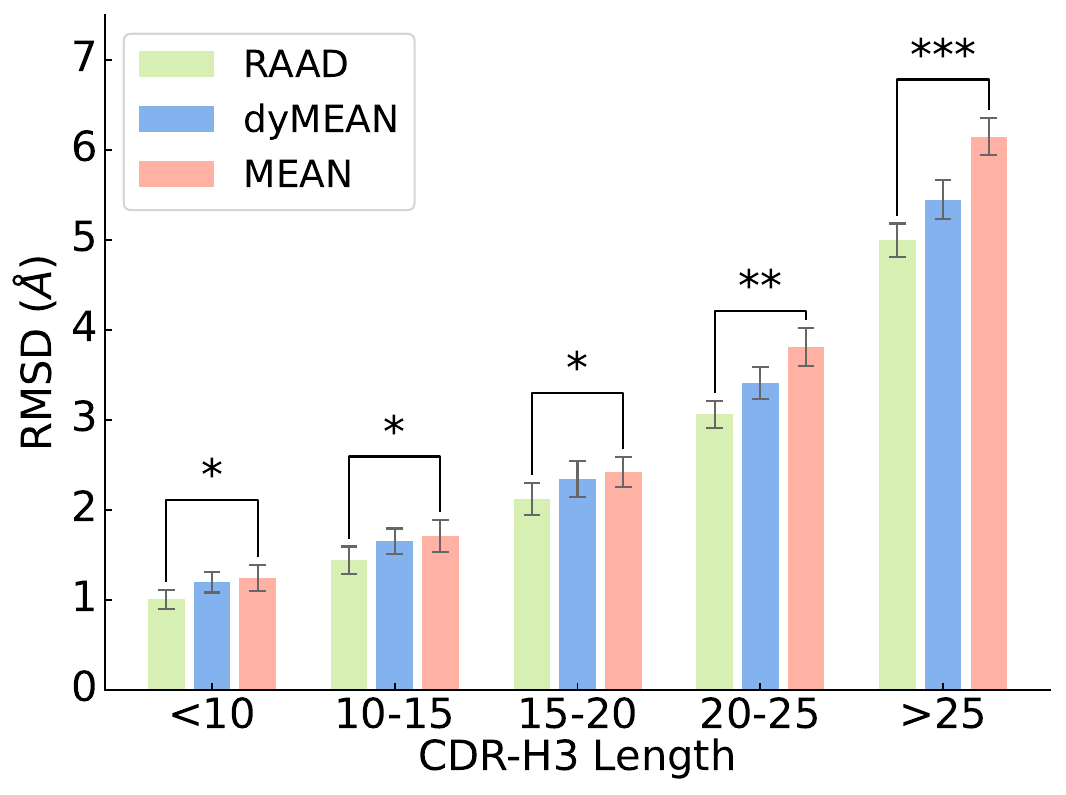}
		\includegraphics[width=0.48\linewidth]{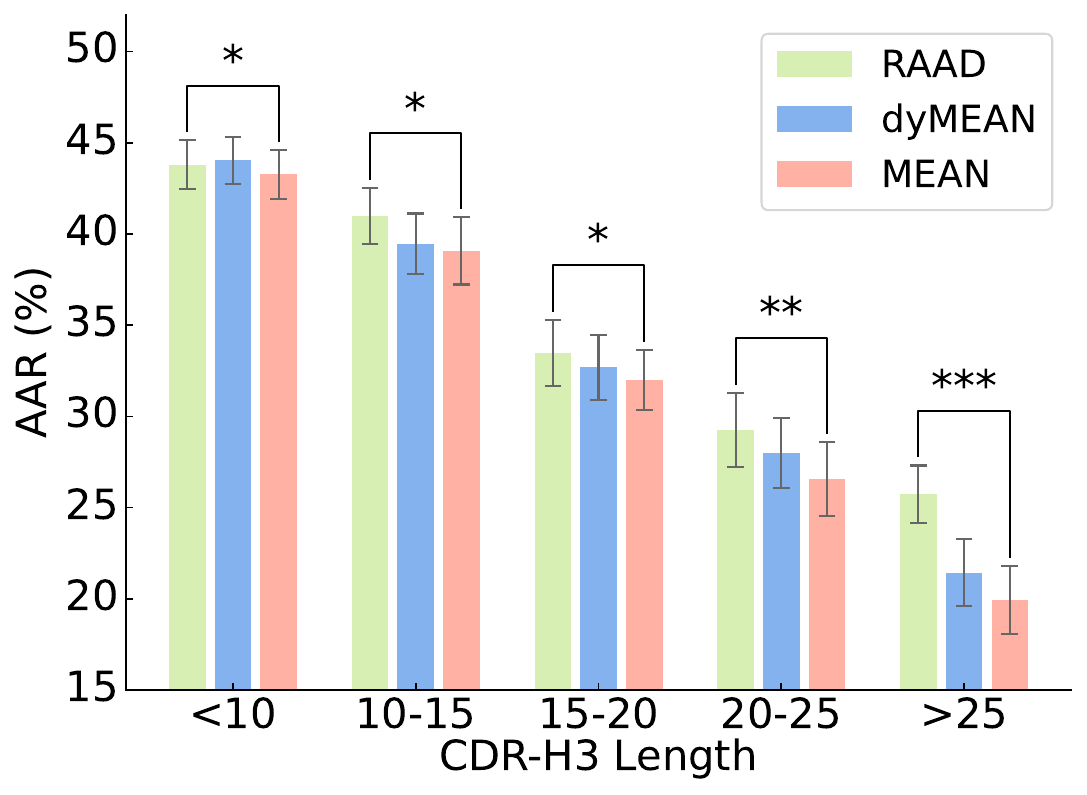}
	\end{center}
	\caption{Performance comparison of three methods for sequence modeling (AAR) and structural modeling (RMSD) for CDR-H3 of different lengths on the SAbDab dataset.}
	\label{fig:8}
\end{figure}

\textbf{CDR Sequence Length.}
To show the superiority of RAAD in handling long CDR sequences, we compare the performance of RAAD with MEAN and dyMEAN under different CDR sequence lengths, including CDR-H3 sequence generation (AAR) and structure prediction (RMSD). As can be seen from the results in Fig.~\ref{fig:8}, the performance gains of RAAD over other baselines keep expanding as the CDR-H3 length increases. This indicates that RAAD has a great advantage in dealing with complex CDR-H3 with long sequences because it captures more contextual information.

\textbf{Pre-training.}
To explore how pre-training influences RAAD in antibody design and optimization, we consider 4 different pre-training schemes. Due to space limitations, a detailed description of these pre-training protocols, experimental results, and analysis can be found in \textbf{Appendix J}.

\section{Conclusion}
In this paper, we identify insufficient contextual information as a major bottleneck in limiting antibody CDR design. To address this problem, we take into account a variety of node and edge features and define multiple relations to achieve \emph{\underline{R}elation-\underline{A}ware \underline{A}ntibody \underline{D}esign} (RAAD) in a one-shot manner. The RAAD framework takes the complete antigen as input and formulates antigen-antibody interaction dynamics modeling as a learnable structural (relational) optimization problem. Furthermore, we propose a new evaluation metric and contrastive constraints for antibody specificity optimization. Extensive experiments are provided to demonstrate the superiority of RAAD in antibody CDR generation and optimization. Despite great progress, limitations still exist in two main aspects: (1) The antibody optimization of our paper relies on an iterative augmentation (ITA) algorithm, where a well-built $\Delta\Delta G$ predictor plays an important role. Combining RAAD with a stronger $\Delta\Delta G$ predictor may further boost the performance. (2) Due to space limitations, only preliminary results on how pre-training affects antibody generation are provided. Combining more powerful protein or antibody pre-training strategies may be beneficial for both antibody generation and optimization.

\section{Acknowledgments}
This work was supported by National Science and Technology Major Project (No. 2022ZD0115101), National Natural Science Foundation of China Project (No. 624B2115, No. U21A20427), Project (No. WU2022A009) from the Center of Synthetic Biology and Integrated Bioengineering of Westlake University and Integrated, and Project (No. WU2023C019) from the Westlake University Industries of the Future Research Funding.

\bibliography{aaai25}

\clearpage
\renewcommand\thefigure{A\arabic{figure}}
\renewcommand\thetable{A\arabic{table}}
\renewcommand\theequation{A.\arabic{equation}}
\setcounter{table}{0}
\setcounter{figure}{0}
\setcounter{theorem}{0}
\setcounter{equation}{0}

\begin{table*}[!htbp]
\begin{center}
\caption{A comparison of various antibody design methods in three aspects: (1) Contextual data, where \textit{``FR"} refers to \textit{``Framework Region"}; (2) Architecture design, where \textit{``Geometry"} refers to the scale of the involved geometric information; and (3) Optimization strategy, where \textit{``Decoding"} refers to whether CDRs are generated in an autoregressive or full-shot fashion, \textit{``No Iterative"} refers to whether iterative refinement or diffusion is used, and \textit{``Pre-train"} refers to the effects of pre-training.}
\label{tab:A1}
\resizebox{\textwidth}{!}{

\begin{tabular}{l|ccc|cc|ccc}
\toprule
\multirow{2}{*}{\textbf{Method}} & \multicolumn{3}{c}{\textbf{Context}} & \multicolumn{2}{c}{\textbf{Architecture}} & \multicolumn{3}{c}{\textbf{Optimization}} \\ \cmidrule(r){2-4} \cmidrule(r){5-6} \cmidrule(r){7-9} 
 & \textbf{\ \ FR\ \ } & \textbf{Light Chain} & \textbf{Antigen} & \textbf{Geometry} & \textbf{Equivariant} & \textbf{Decoding} & \textbf{No Iterative} & \textbf{Pre-train} \\ \midrule
RefineGNN (ICLR'22) & \Checkmark & \XSolidBrush & \XSolidBrush & Backbone & \XSolidBrush & Autoregressive & \XSolidBrush & \XSolidBrush \\
HSRN  (ICML'22) & \XSolidBrush & \XSolidBrush & Epitope & Full-atom & \Checkmark & Autoregressive & \XSolidBrush & \XSolidBrush \\
DiffAb  (NIPS'22) & \Checkmark & \XSolidBrush & Full & Residue (C-$\alpha$) & \Checkmark & Full-shot & \XSolidBrush & \XSolidBrush \\
MEAN  (ICLR'23) & \Checkmark & \Checkmark & Epitope & Backbone & \Checkmark & Full-shot & \XSolidBrush & \XSolidBrush \\
dyMEAN  (ICML'23) & \Checkmark & \Checkmark & Epitope & Full-atom & \Checkmark & Full-shot & \XSolidBrush & \XSolidBrush \\
ABGNN  (KDD'23) & \Checkmark & \XSolidBrush & Epitope & Full-atom & \Checkmark & Full-shot & \XSolidBrush & \Checkmark \\
HTP  (NeurIPS'23) & \Checkmark & \Checkmark & Epitope & Residue (C-$\alpha$) & \Checkmark & Full-shot & \Checkmark & \Checkmark \\
ADsigner  (AAAI'24) & \Checkmark & \Checkmark & Epitope & Backbone & \XSolidBrush & Full-shot & \Checkmark & \XSolidBrush \\
RAAD (ours) & \Checkmark & \Checkmark & Full & Backbone & \Checkmark & Full-shot & \Checkmark & \Checkmark \\ \bottomrule
\end{tabular} 

}
\end{center}
\end{table*} 

\section*{Appendix}

\subsection*{A. Comparison of Method Characteristics} 
We compare various methods in Table.~\ref{tab:A1} from three key aspects: \textit{(1) input contextual data}, including framework regions on the heavy chain, light chain, and antigen; \textit{(2) architectural design}, the modeling levels of geometric information, and whether the modules are E(3)-equivariant; and \textit{(3) optimization strategy}, whether it is autoregressive, whether it is iterative refinement-based or diffusion-based, and whether it evaluates the effects of model pre-training.

\subsection*{B. Definitions of Node and Edge Features}
\begin{figure}[!htbp]
\vspace{-0.5em}
\begin{center}\includegraphics[width=0.9\linewidth]{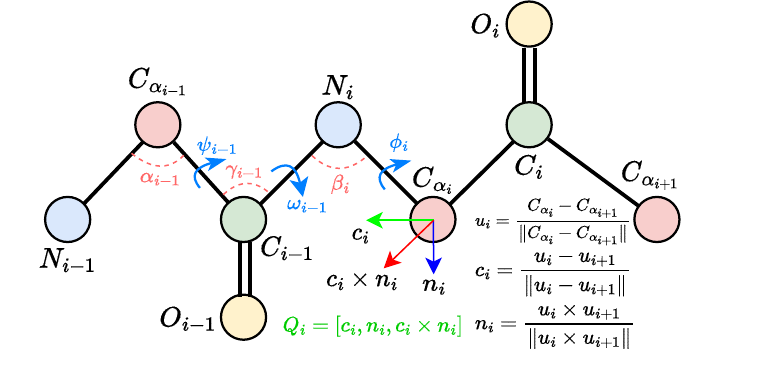}
	\end{center}

    \resizebox{\columnwidth}{!}{
    \renewcommand\arraystretch{1.2}
    \begin{tabular}{l|l|l}
    \toprule
     & \multicolumn{1}{c|}{Node $v_i$} & \multicolumn{1}{c}{Edge $e_{i,j}$} \\ \hline
    Type & $E_{\text{type}}(v_i)$ & $E_{\text{type}}(e_{i,j})$ \\
    Position & $E_{\text{pos}}(i)$ & $E_{\text{pos}}(i-j)$ \\
    Distance & \tabincell{l}{$\left\|C_{\alpha_i}-C_i\right\|$\ \  $\left\|C_{\alpha_i}-N_i\right\|$ \\ $\left\|C_{\alpha_i}-O_i\right\|$} & \tabincell{l}{$\left\|C_{\alpha_i}-C_j\right\|$ \ \ $\left\|C_{\alpha_i}-N_j\right\|$ \\ $\left\|C_{\alpha_i}-O_j\right\|$ \ \ $\left\|C_{\alpha_i}-C_{\alpha_j}\right\|$} \\
    Angle & \tabincell{l}{Angle - $\alpha_i$, $\beta_i$, $\gamma_i$ \\ Dihedral Angle - $\phi_i$, $\psi_i$, $\omega_i$} & \multicolumn{1}{c}{—} \\
    Direction & \tabincell{l}{$Q_i^{\top}\frac{C_i-C_{\alpha_i}}{\left\|C_i-C_{\alpha_i}\right\|}$\ \  $Q_i^{\top}\frac{N_i-C_{\alpha_i}}{\left\|N_i-C_{\alpha_i}\right\|}$ \\ $Q_i^{\top}\frac{O_i-C_{\alpha_i}}{\left\|O_i-C_{\alpha_i}\right\|}$} & \tabincell{l}{$Q_i^{\top}\frac{C_j-C_{\alpha_i}}{\left\|C_j-C_{\alpha_i}\right\|}$\ \ $Q_i^{\top}\frac{N_j-C_{\alpha_i}}{\left\|N_j-C_{\alpha_i}\right\|}$ \\ $Q_i^{\top}\frac{O_j-C_{\alpha_i}}{\left\|O_j-C_{\alpha_i}\right\|}$\ \ $Q_i^{\top}\frac{C_{\alpha_j}-C_{\alpha_i}}{\left\|C_{\alpha_j}-C_{\alpha_i}\right\|}$} \\
    Orientation & \multicolumn{1}{c|}{—} & \multicolumn{1}{c}{$q\left(Q_i^{\top} Q_j\right)$} \\ \bottomrule
    \end{tabular}}

	\caption{Node and edge features for single or pairwise residues, all invariant to rotation and translation. The definitions of bond angles ($\alpha_i$, $\beta_i$, $\gamma$), dihedral angles ($\psi_i$, $\phi_i$, $\omega_i$), and local coordinates frame $Q_i$ refer to~\cite{gao2022pifold}.}
	\label{fig:A1}
\end{figure}

The node and edge features are defined in terms of six key aspects: type, position, distance, direction, angle, and orientation. A summary of these features is available in Fig.~\ref{fig:A1}.

\subsection*{C. Illustration of Edge Relations} 
\begin{figure}[!htbp]
    \vspace{-0.5em}
    \begin{center}
	\includegraphics[width=0.8\linewidth]{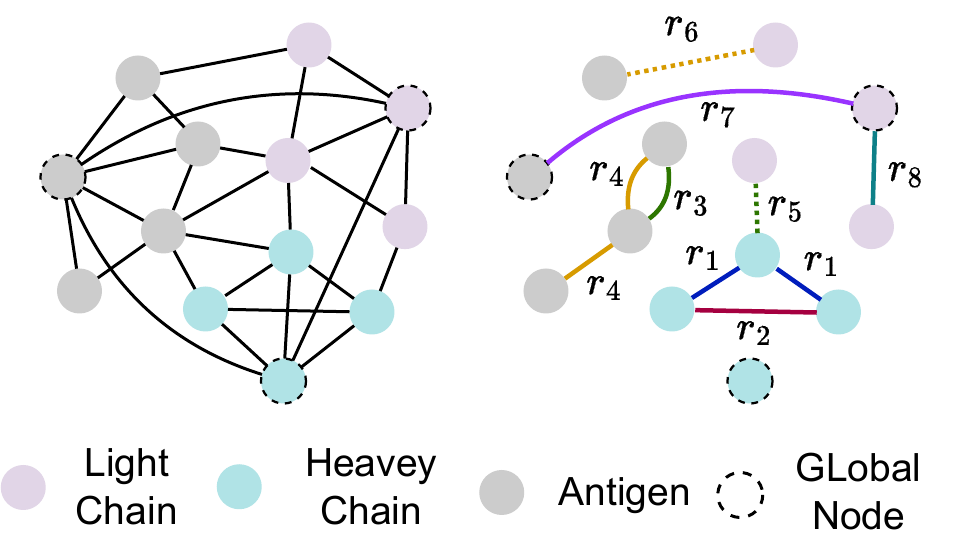}
    \end{center}
    \vspace{-1em}
    \caption{Illustration of edges relations. To avoid overly complexity, we visualize only some edges.}
    \vspace{-1em}
    \label{fig:A2}
\end{figure}

We provide a schematic of edges relations in Fig.~\ref{fig:A2}, where each edge $e_{i,j}\in\mathcal{E}$ is associated with a set of relations $R_{i,j}\in\mathcal{R}$. Besides, two relations $r_1$ (with sequence distance equal to 1) can derive a relation $r_2$ (with sequence distance equal to 2). In addition, an edge may connect two nodes (residues) due to both relations $r_3$ and $r_4$. We strongly encourage readers to combine Fig.~\ref{fig:A2} with the descriptions in the paper to understand the definition of edge relations.

\subsection*{D. Feature/Coordinate Updating of Global Nodes} 
While node features and coordinates of global nodes can be updated as defined in Eq.~(\ref{equ:5})(\ref{equ:7}), we further provide their detailed derivations as below. The node feature of the global node $g_k$ $(1\leq k \leq 3)$ of the $k$-th context component can be updated by local and global interactions, as follows:
\begin{equation}
\begin{aligned}
\mathbf{h}_{g_k}^{(l+1)}\!=\!\phi_h \Big(\mathbf{h}_{g_k}^{(l)}, \underbrace{\boldsymbol{W}_{r_7}^{(l)}\!\! \sum_{g_l\neq g_k} \alpha_{g_k,g_l}^{(r_7)} \ \mathbf{m}_{g_k,g_l,r_7}^{(l)}}_{\text{global interaction}} \\ + \underbrace{\boldsymbol{W}_{r_8}^{(l)}\!\! \sum_{\Phi(j)=g_k} \alpha_{g_k,j}^{(r_8)} \ \mathbf{m}_{g_k,j,r_8}^{(l)}}_{\text{local interaction}}\Big)
\end{aligned}
\end{equation}
where $\alpha_{g_k,g_l,r_7}^{(l)}$ and $\alpha_{g_k,j,r_8}^{(l)}$ are attention weights as calculated by Eq.~(\ref{equ:5}). Besides, $\Phi(j)=g_k$ is the indicator function that determines whether node $v_j\in\mathcal{V}$ and $v_{g_k}$ belong to the same component. The node coordinate of the global node $g_k$ is updated by local and global interactions, as follows:
\begin{equation}
\hspace{-1.0em}
\begin{aligned}
\mathbf{X}_{g_k}^{(l+1)}&\!=\!\mathbf{X}_{g_k}^{(l)}\!+\!\frac{1}{2}\Big(\underbrace{\sum_{g_l\neq g_k} \alpha_{g_k,g_l}^{(r_7)} \mathbf{X}_{g_k,g_l}^{(l)} \phi_z^{(r_7)}\big(\mathbf{m}_{g_k,g_l,r_7}^{(l)}\big)}_{\text{global interaction}} \\ & + \underbrace{\sum_{\Phi(j)=g_k}\alpha_{g_k,j}^{(r_8)} \mathbf{X}_{g_k,j}^{(l)} \phi_z^{(r_8)}\big(\mathbf{m}_{g_k,j,r_8}^{(l)}\big)}_{\text{local interaction}} \Big) 
\end{aligned}
\end{equation}

\vspace{-1em}
\subsection*{E. On the proof of RA-EGN E(3)-Equivariance} 
\begin{theorem}
We denote the $l$-th RA-EGN layer $g^{(l)}(\cdot)$ in our framework as $(\mathbf{H}^{(l+1)}, \mathbf{X}^{(l+1)}, \mathbf{E}^{(l+1)}, R^{(l+1)})=g^{(l)}(\mathbf{H}^{(l)}, \mathbf{X}^{(l)}, \mathbf{E}^{(l)}, R^{(l)})$. For any given orthogonal matrix $\mathbf{O}\in\mathbb{R}^{3\times 3}$ and translation matrix $\mathbf{t}\in\mathbb{R}^{3}$, $g^{(l)}(\cdot)$ is E(3)-equivariant, i.e., we have $(\mathbf{H}^{(l+1)}, \mathbf{O}\mathbf{X}^{(l+1)} + \mathbf{t}, \mathbf{E}^{(l+1)}, R^{(l+1)})=g^{(l)}(\mathbf{H}^{(l)}, \mathbf{O}\mathbf{X}^{(l)}+\mathbf{t}, \mathbf{E}^{(l)}, R^{(l)})$.
\end{theorem}

\textit{Proof.}
For any orthogonal matrix $\mathbf{O}\in\mathbb{R}^{3\times 3}$, translation matrix $\mathbf{t}\in\mathbb{R}^{3}$, and $\mathbf{X}_{i,j}^{(l)}=\mathbf{X}_{i}^{(l)}-\mathbf{X}_{j}^{(l)}$, We can derive the relative coordinates between $\mathbf{O}\mathbf{X}_{i}^{(l)}+\mathbf{t}$ and $\mathbf{O}\mathbf{X}_{j}^{(l)}+\mathbf{t}$ as
\begin{equation}
    \hspace{-1em}
    (\mathbf{O}\mathbf{X}_{i}^{(l)}+\mathbf{t}) \!-\! (\mathbf{O}\mathbf{X}_{j}^{(l)}+\mathbf{t})\!=\!\mathbf{O}\mathbf{X}_{i}^{(l)}-\mathbf{X}_{j}^{(l)})\!=\!\mathbf{O}\mathbf{X}_{i,j}^{(l)}
\end{equation}
The message aggregation of Eq.~(\ref{equ:3}) can be re-writed as
\begin{equation}
\hspace{-1em}
\begin{aligned}
&\phi_m\Big(\mathbf{h}_i^{(l)}, \mathbf{h}_j^{(l)}, \frac{(\mathbf{O}\mathbf{X}_{i,j}^{(l)})^{\top} \mathbf{O}\mathbf{X}_{i,j}^{(l)}}{\big\|(\mathbf{O}\mathbf{X}_{i,j}^{(l)})^{\top} \mathbf{O}\mathbf{X}_{i,j}^{(l)}\big\|_F}, \mathbf{E}_{i,j}^{(l)}\Big) \\
&=\phi_m\Big(\mathbf{h}_i^{(l)}, \mathbf{h}_j^{(l)}, \frac{(\mathbf{X}_{i,j}^{(l)})^{\top}\mathbf{O}^{\top} \mathbf{O}\mathbf{X}_{i,j}^{(l)}}{\big\|(\mathbf{X}_{i,j}^{(l)})^{\top}\mathbf{O}^{\top} \mathbf{O}\mathbf{X}_{i,j}^{(l)}\big\|_F}, \mathbf{E}_{i,j}^{(l)}\Big) \\
&=\phi_m\Big(\mathbf{h}_i^{(l)}, \mathbf{h}_j^{(l)}, \frac{(\mathbf{X}_{i,j}^{(l)})^{\top} \mathbf{X}_{i,j}^{(l)}}{\big\|(\mathbf{X}_{i,j}^{(l)})^{\top} \mathbf{X}_{i,j}^{(l)}\big\|_F}, \mathbf{E}_{i,j}^{(l)}\Big)  = \mathbf{m}_{i,j,r}^{(l)}
\end{aligned}
\end{equation}
Therefore, the message aggregation is $E(3)$-invariant. Since the input node feature $\mathbf{h}^{(l)}_i$, edge feature $\mathbf{E}^{(l)}_{i,j}$, and message $\mathbf{m}_{i,j,r}^{(l)}$ are all $E(3)$-invariant, it is easy to derive that updated node features, edge features and edge relations in the $l$-th RA-EDGN layer are also $E(3)$-invariant. 

Following that, we can prove that the node coordinates are also $E(3)$-
equivariant, as follows:
\begin{equation}
\begin{aligned}
&\big(\mathbf{O}\mathbf{X}_i^{(l)} + \mathbf{t} \big)\!+\!\frac{1}{|\mathcal{R}|} \sum_{r\in\mathcal{R}} \sum_{j \in \mathcal{N}_i^{(r)}} \!\alpha_{i,j,r}^{(l)}\ \big(\mathbf{O}\mathbf{X}_{i,j}^{(l)}\big)\ \phi_z^{(r)}\big(\mathbf{m}_{i,j,r}^{(l)}\big) \\
& = \mathbf{O}\Big(\mathbf{X}_i^{(l)} \!+\!\frac{1}{|\mathcal{R}|} \sum_{r\in\mathcal{R}} \sum_{j \in \mathcal{N}_i^{(r)}} \!\alpha_{i,j,r}^{(l)}\ \mathbf{X}_{i,j}^{(l)}\ \phi_z^{(r)}\big(\mathbf{m}_{i,j,r}^{(l)}\big)\Big)+ \mathbf{t} \\
& = \mathbf{O}\mathbf{X}_i^{(l+1)}+ \mathbf{t}
\end{aligned}
\end{equation}

With the above derivations, it is easy to derive that
\begin{equation}
\begin{aligned}
    &(\mathbf{H}^{(l+1)}, \mathbf{O}\mathbf{X}^{(l+1)} + \mathbf{t}, \mathbf{E}^{(l+1)}, R^{(l+1)}) \\ &=g^{(l)}(\mathbf{H}^{(l)}, \mathbf{O}\mathbf{X}^{(l)}+\mathbf{t}, \mathbf{E}^{(l)}, R^{(l)})
\end{aligned}
\end{equation}
For an $L$-layer RA-EGN encoder, we have
\begin{equation}
\begin{aligned}
    &(\mathbf{H}^{(L)}, \mathbf{O}\mathbf{X}^{(L)} + \mathbf{t}, \mathbf{E}^{(L)}, R^{(L)}) \\ &
    = g^{(L-1)}(\mathbf{H}^{(L-1)}, \mathbf{O}\mathbf{X}^{(L-1)}+\mathbf{t}, \mathbf{E}^{(L-1)}, R^{(L-1)}) \\ &
    = g^{(L-2)}(\mathbf{H}^{(L-2)}, \mathbf{O}\mathbf{X}^{(L-2)}+\mathbf{t}, \mathbf{E}^{(L-2)}, R^{(L-2)}) \\ &
    = \cdots \\ &
    = g^{(0)}(\mathbf{H}^{(0)}, \mathbf{O}\mathbf{X}^{(0)}+\mathbf{t}, \mathbf{E}^{(0)}, R^{(0)})
\end{aligned}
\end{equation}
Since $\mathbf{H}^{(0)}$, $\mathbf{E}^{(0)}$, and $R^{(0)}$ are constructed based on the backbone coordinates $\mathbf{X}^{(0)}$ and keep $E(3)$-invariant, the resulting encoder is also $E(3)$-equivariant as a whole.

\begin{figure}[!tbp]
    \begin{center}  \includegraphics[width=1.0\linewidth]{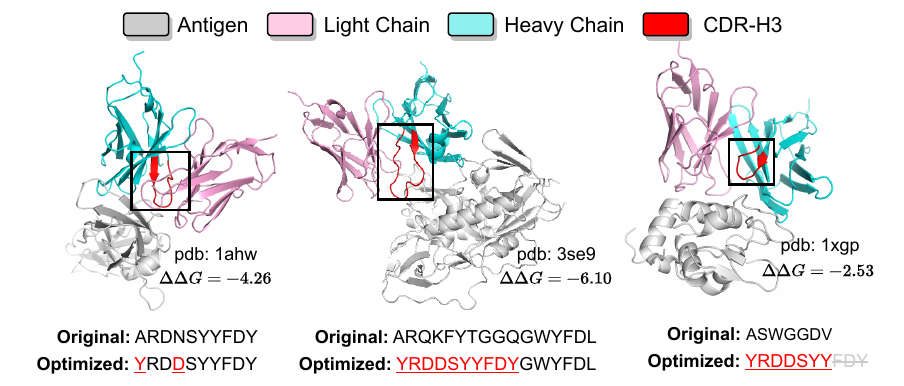}
    \end{center}
	\caption{A case study of ``universal" antibodies, in which optimized CDR sequences tailored for one antigen (pdb: 1ahw) can be used for other antigens (pdb: 3se9 and 1xgp) with some minor length adaptations (e.g., replacements and cropping) to improve their binding affinity, i.e., lower $\Delta\Delta G$.}
    \label{fig:A3}
\end{figure}

\subsection*{F. Huber Loss Function} 
The huber loss~\cite{huber1992robust} helps to lead to a stable training procedure~\cite{kong2022conditional}, defined as follows:
\begin{equation}
l(x, y)=\left\{\begin{array}{l}
0.5(x-y)^2, \text { if }|x-y|<\delta \\
\delta \cdot(|x-y|-0.5 \cdot \delta), \text { else }
\end{array}\right.
\end{equation}
where we set $\delta=1$ in this paper.

\begin{table*}[!tbp]
\begin{center}
\caption{Performance comparison of models pre-trained with different data, including 1D protein sequences (UniRef), 1D antibody sequences (OAS), 3D protein structures (AlphaFoldDB), as well as hierarchical pre-training.}
\label{tab:A2}
\resizebox{1.0\textwidth}{!}{
\begin{tabular}{cc|cccccc}

\toprule
\multicolumn{2}{c}{\textbf{Pre-training}} & \multicolumn{2}{c}{\textbf{Modeling}} & \multicolumn{3}{c}{\textbf{Generation}} & \textbf{Optimization} \\ \cmidrule(r){1-2} \cmidrule(r){3-4} \cmidrule(r){5-7} \cmidrule(r){8-8}   
Encoder & Pre-training Data & AAR ($\uparrow$) & RMSD ($\downarrow$) & AAR ($\uparrow$) & TM-score ($\uparrow$) & RMSD ($\downarrow$) & SP-score ($\uparrow$) \\ \midrule
\multicolumn{2}{c|}{no pre-training} & 67.64 & 0.772 & 41.26 & 0.9874 & 1.463 & \underline{11.55} \\
ESM-1b & UniRef & 68.10 & 0.768 & 41.40 & 0.9866 & 1.475 & 10.94 \\
AbBERT & OAS & \underline{68.32} & \textbf{0.738} & \underline{41.78} & 0.9876 & 1.437 & 11.37 \\
GearNet-Edge & AlphaFoldDB & 66.38 & 0.761 & 40.19 & \underline{0.9883} & \underline{1.413} & \textbf{11.87} \\
\multicolumn{1}{c}{ESM-1b} & UniRef \& OAS & \textbf{68.74} & \underline{0.745} & \textbf{41.94} & \textbf{0.9890} & \textbf{1.394} & 11.17 \\ \bottomrule

\end{tabular}}
\end{center}
\end{table*}

\subsection*{G. Iterative Target Augmentation (ITA)} 
The pseudo-code of ITA-based antibody specificity optimization is placed in Algorithm.~\ref{algo:1}.

\begin{algorithm}[!htbp]
    \caption{Algorithm for ITA-based Optimization}
    \label{algo:1}
    \begin{algorithmic}[1]
  
    \Require  A set of antibodies $\mathcal{D}$ to be optimized, a antibody generator $\mathcal{F}_{\Theta}$, and a $\Delta\Delta G$ predictor $f$.

    \For{each iteration}
        \For{each antibody $s$ in $\mathcal{D}$}
            \State Remove CDRs from antibody $s$ and generate 50 candidate antibodies by generator $\mathcal{F}_{\Theta}$.
            \State  Sort candidates by predictor $f$, update the candidate queue, and update the training set $\mathcal{Q}$. 
        \EndFor
        \State Sample a batch of high-affinity antibodies from $\mathcal{Q}$.
        \State Fine-tune the generator $\mathcal{F}_{\Theta}$ by minimizing the total loss $\mathcal{L}_{\text{opt}}$ in Eq.~(\ref{equ:12}) by back-propagation.
	\EndFor
	
    Generate high-affinity antibodies with generator $\mathcal{F}_{\Theta_*}$.
        
	\State \textbf{return} Fine-tuned antibody generator $\mathcal{F}_{\Theta_*}$.
    \end{algorithmic}
\end{algorithm}

\subsection*{H. Time Complexity Analysis}
The time complexity of the relation-aware encoder (consisting of multiple layers of RA-EGNs) comes from five main components: (1) message aggregation $\mathcal{O}\big(|\mathcal{E}|(d_nF+d_eF)\big)$
, where $|\mathcal{E}|$ is the number of edges, $d_n$ and $d_e$ are feature dimension of nodes and edges, and $F$ is the hidden dimension;  (2) node relation updating $\mathcal{O}(|\mathcal{E}|F)$; (3) node feature updating $\mathcal{O}\big(|\mathcal{R}|\cdot|\mathcal{V}|F^2+ |\mathcal{R}|\cdot|\mathcal{E}|(d_n+F)\big)$, where $|\mathcal{V}|$ and $|\mathcal{R}|$ is the number of nodes and relations; (4) edge feature updating $\mathcal{O}\big(|\mathcal{E}|(d_nF+d_eF)\big)$; (5) node coordinate updating $\mathcal{O}(|\mathcal{R}|\cdot|\mathcal{E}|F)$. The total time complexity is $\mathcal{O}\Big(|\mathcal{E}|(d_n+d_e)F+|\mathcal{R}|\cdot\big(|\mathcal{V}|F^2+|\mathcal{E}|(d_n+F)\big)\Big)$ is linear w.r.t the number of nodes $|\mathcal{V}|$ and edges $|\mathcal{E}|$. In this paper, we define a total of 8 edge relations, i.e., $|\mathcal{R}|\!=\!8\!\ll\!|\mathcal{V}|$, which is almost negligible in the time complexity analysis.

\subsection*{I. Implementation Details and Hyperparameters}
The following hyperparameters are determined by an AutoML toolkit NNI with the hyperparameter search spaces as:  Adam optimizer~\cite{kingma2014adam} with $lr=\{0.0005, 0.001\}$, batch size $B=\{4, 8, 16\}$, training epoch $E_{\text{train}}=20$, fine-tuning epoch $E_{\text{fine}}=20$, number of nearest neighbors $K=8$ for graph construction, embedding size $F_e=32$ for $E_{\text{type}}(v_i)$, hidden dimension $F=\{128, 256\}$, layer number $L=\{3, 4\}$, and loss weight $\eta=\{0.8, 1.0\}$ for $\mathcal{L}_{\text{struct}}$. In addition, $\phi_m(\cdot)$, $\phi_z^{(r)}(\cdot)$, $\phi_{\omega}^{(r)}(\cdot)$, $\phi_h(\cdot)$, and $\phi_e(\cdot)$ are all implemented as one- or two-layer MLPs with $\operatorname{SiLU}(\cdot)$~\cite{elfwing2018sigmoid} as the activation function in this paper. For a fairer comparison, we select the mode checkpoint with the lowest loss on the validation set for testing. Moreover, the experiments on both baselines and our approach are implemented based on the standard implementation using the PyTorch 1.8.0 with Intel(R) Xeon(R) Gold 6240R @ 2.40GHz CPU and 8 NVIDIA A100 GPUs. For complex initialization, we remove the CDR $\mathcal{C}$ from the antibody, and then we use HDOCK \cite{yan2017hdock} to dock it to the target antigen to obtain an antibody-antigen complex. Since the CDR $\mathcal{C}$ is unknown at first, we initialize its 3D coordinates according to the even distribution between the two residues right before and after the CDR.

\subsection*{J. Effects of Pre-training}
To explore how pre-training influences RAAD in antibody design, we consider four pre-training schemes: (1) ESM-1b~\cite{rives2021biological} encoder pre-trained on the UniRef dataset~\cite{suzek2007uniref}; (2) AbBERT~\cite{gao2023pre} encoder pre-trained on the OAS dataset~\cite{olsen2022observed}; (3) GearNet-edge~\cite{zhang2022protein} pre-trained on the AlphaFoldDB dataset~\cite{varadi2022alphafold}; (4) ESM-1b encoder hierarchically pre-trained on the UniRef and OAS datasets \cite{wu2023hierarchical}. Following~\cite{gao2023pre}, we exploit the pre-trained encoders by concatenating the pre-trained and original node features, forwarding with a linear transformation, and taking the fused features as inputs. It can be observed from Table.~\ref{tab:A2} that (1) all kinds of pre-training strategies help to improve performance; (2) sequence pre-training helps more for sequence generation, while structure pre-training is more beneficial for structure prediction; (3) pre-training with antibody sequences helps much more than general protein sequences, but combining the two for hierarchical pre-training outperforms the both.

\end{document}